\documentclass{aims} 
\usepackage{amsmath}
\usepackage{paralist}
\usepackage[misc]{ifsym}
\usepackage{epsfig} 
\usepackage{epstopdf} 
\usepackage[colorlinks=true]{hyperref}
\hypersetup{urlcolor=blue, citecolor=red}
\allowdisplaybreaks

\def\currentvolume{X}
 \def\currentissue{X}
  \def\currentyear{200X}
   \def\currentmonth{XX}
    \def\ppages{X-XX}
     \def\DOI{10.3934/xx.xxxxxxx}
     
\usepackage[capitalize]{cleveref}
\usepackage{siunitx}
\usepackage{booktabs}
\usepackage{bm}
\usepackage{caption}
\usepackage{todonotes}
\usepackage[acronym]{glossaries}
\usepackage{wrapfig}
\usepackage{subfig}
\usepackage{enumitem}

\theoremstyle{definition}

\crefname{enumi}{}{}

\newcommand{\eqdot}{\quad \text{.}}
\newcommand{\eqcolon}{\quad \text{,}}
\newcommand{\etal}{et al.\ }
\newacronym{CAD}{CAD}{Computer Aided Design}
\newacronym{CFD}{CFD}{Computational Fluid Dynamics}
\newacronym{DOF}{DOF}{Degree of Freedom}
\newacronym{DQN}{DQN}{Deep Q-Network}
\newacronym{DRL}{DRL}{Deep Reinforcement Learning}
\newacronym{FEM}{FEM}{Finite Element Method}
\newacronym{FFD}{FFD}{Free Form Deformation}
\newacronym{NN}{NN}{Neural Network}
\newacronym{NURBS}{NURBS}{Non-Rational Uniform B-Splines}
\newacronym{PDE}{PDE}{Partial Differential Equation}
\newacronym{PPO}{PPO}{Proximal Policy Optimization}
\newacronym{RL}{RL}{Reinforcement Learning}
\newacronym{SAC}{SAC}{Soft Actor Critic}
\newacronym{DDPG}{DDPG}{Deep Deterministic Policy Gradient}
\newacronym{A2C}{A2C}{Advantage Actor Critic}

\title[]
{Investigation of reinforcement learning for shape optimization of profile extrusion dies} 

\author[]{}



\thanks{$^*$Corresponding author: Daniel Wolff}

\begin{document}
\maketitle

\centerline{\scshape
Clemens Fricke$^{{\href{mailto:clemens.david.fricke@tuwien.ac.at}{\textrm{\Letter}}}1}$
Daniel Wolff$^{{\href{mailto:wolff@cats.rwth-aachen.de}{\textrm{\Letter}}}*2}$
Marco Kemmerling$^{{\href{mailto:marco.kemmerling@ima.rwth-aachen.de}{\textrm{\Letter}}}3}$
and Stefanie Elgeti$^{{\href{mailto:stefanie.elgeti@tuwien.ac.at}{\textrm{\Letter}}}1}$}

\medskip

{\footnotesize
 \centerline{$^1$Institute for Lightweight Design and Structural Biomechanics, TU Wien, Austria}
} 

\medskip

{\footnotesize
 \centerline{$^2$Chair for Computational Analysis of Technical Systems, RWTH Aachen University, Germany}
}

\medskip

{\footnotesize
 \centerline{$^2$Information Management in Mechanical Engineering, RWTH Aachen University, Germany}
}

\bigskip

 \centerline{(Communicated by Handling Editor)}


\begin{abstract}
Profile extrusion is a continuous production process for manufacturing plastic profiles from molten polymer. Especially interesting is the design of the die, through which the melt is pressed to attain the desired shape. However, due to an inhomogeneous velocity distribution at the die exit or residual stresses inside the extrudate, the final shape of the manufactured part often deviates from the desired one. To avoid these deviations, the shape of the die can be computationally optimized, which has already been investigated in the literature using classical optimization approaches \cite{Elgeti2012,Rajkumar2018,Zhang2019a}. 

A new approach in the field of shape optimization is the utilization of \gls{RL} as a learning-based optimization algorithm. \gls{RL} is based on trial-and-error interactions of an agent with an environment. For each action, the agent is rewarded and informed about the subsequent state of the environment. While not necessarily superior to classical, e.g., gradient-based or evolutionary, optimization algorithms for one single problem, \gls{RL} techniques are expected to perform especially well when similar optimization tasks are repeated since the agent learns a more general strategy for generating optimal shapes instead of concentrating on just one single problem.

In this work, we investigate this approach by applying it to two 2D test cases.
The flow-channel geometry can be modified by the \gls{RL} agent using so-called \acrlong{FFD} \cite{Sederberg1986}, a method where the computational mesh is embedded into a transformation spline, which is then manipulated based on the control-point positions.  
In particular, we investigate the impact of utilizing different agents on the training progress and the potential of wall time saving by utilizing multiple environments during training.
\end{abstract}


\section{Introduction to Profile Extrusion and Reinforcement Learning}
\label{sec:Introduction}

Profile extrusion is a widely used and -- by now -- well-established continuous production process for manufacturing plastic profiles with a desired cross-sectional shape. Its application ranges from fabricating plastic films and sealings via cable channels to floor skirtings and window frames. Despite its vast application, the process is connected to a couple of challenges that need to be overcome in order to run it efficiently. So far, the design process of extrusion dies for new cross-section shapes was mainly guided by the experience of the engineer in charge. However, designing flow channels to distribute the flow of the highly-viscous plastic melt inside a profile extrusion die is not a trivial task. This is due to its strongly non-linear rheological behavior, referred to as \textit{shear-thinning}  \cite{Osswald2015}. Inhomogeneous velocity distributions at the die's outlet and remaining residual stresses inside the extruded, solidified part may result in significant deviations from the desired target shape, known as \textit{warpage}. Consequently, homogeneous velocities at the outlet are considered the most relevant criterion for designing these flow channels \cite{Nobrega2004,Hopmann2016}. Moreover, to increase the overall efficiency of the production process, it is of great interest to reduce material consumption during manufacturing, e.g., by reducing the amount of scrap. Here, the computational design of flow channels inside profile extrusion dies comes in handy, as it allows the numerical investigation of the design criterion (e.g., a homogeneous velocity distribution at the extruder outlet) and thus generates a design that is close to optimal with respect to the chosen design criterion.

To optimize the design of the flow channels, different design criteria can be taken into consideration. These include e.g. a low-pressure drop, short residence times, or uniform die swells, while the most influential quantity with respect to the quality of the manufactured part is the flow homogeneity at the die exit, i.e., a uniform velocity distribution is desired \cite{Hopmann2016}. The (shape) optimization of profile extrusion dies has been an active field of research for many years \cite{Elgeti2012,Siegbert2014,Yilmaz2014}. As pointed out in the review paper by Pittmann \cite{Pittman2011}, the initial guidelines for design procedures were mainly driven by previous experiences, pocket calculations, and analytical formulas derived from reasonable simplifications of the governing conservation laws. However, with the increase in computational capabilities, computer-based design approaches have emerged, since they promise to be economically and efficiently beneficial. To optimize the flow channels inside dies computationally, the above-mentioned design criteria need to be translated into definitions of suitable objective functions.

Already in 2000, Szarvasy \etal \cite{Szarvasy2000} proposed an objective function for computational die design, which depends on the local mass-flow rates in different geometric partitions of the profile extrusion die's outflow cross-section. The main purpose of their work, however, was to compare the objective function values computed from numerical \gls{FEM} simulations to those obtained from experimentally measured data. They tested their proposed objective function on T- and Y-shaped profiles and used the channel height of the interior branches as a design variable. The numerical and experimental data proved to be in poor agreement, which was mostly attributed to neglected physical effects within the \gls{FEM} simulations. Especially, properly modeling the rheological behavior of the plastic melt is important to obtain meaningful results. 
Another very early approach presented by Michaeli \etal \cite{Michaeli2001} combined \gls{FEM} and network theory to drastically reduce the number of required iterations in the design process. Their basic idea was that even complex extrusion dies can be represented by a connection of simpler geometric sections. Averaging the results of the \gls{FEM} simulation in each section, a network model was derived, which allowed to relate the objective function, i.e., a uniform velocity distribution at the die outlet, to the flow resistances associated with each section. Based on the optimization of these flow resistances (either manually or by an optimization algorithm), an optimized geometry was obtained. \\
The advances in computational capabilities allowed for direct optimization approaches, which fully relied on computer simulations without the need of leveraging network theory. Elgeti \etal \cite{Elgeti2012} proposed a shape optimization framework for solving the \gls{PDE}-constrained shape optimization problem on different die geometries. To represent the rheological behavior of the molten plastic, a non-linear shear-thinning Carreau model was employed. The resulting nonlinear equation system was then solved by means of the \gls{FEM}. The boundary of the die geometry was represented by \gls{NURBS}, a certain type of spline, whose control point coordinates became the degrees of freedom of the shape optimization. Based on this parameterization, the computational meshes for the modified die geometries were obtained by updating an initial mesh with the help of a mesh-update method. The authors also optimized the die geometry with respect to a homogeneous velocity distribution at the outlet with the help of a derivative-free optimization algorithm. A different objective function was considered by Pauli \etal \cite{Pauli2013} who optimized profile extrusion dies with respect to a homogeneous die swell, taking into consideration the viscoelastic behavior of plastic melts. To parameterize the geometry, the same approach as in \cite{Elgeti2012} was employed alongside the same derivative-free optimization algorithm. Siegbert \etal \cite{Siegbert2014} investigated the influence of different algorithms on the optimization process. \\
Rajkumar \etal \cite{Rajkumar2017} also optimized the extrusion die geometry with respect to a homogeneous velocity distribution, however using a slightly different objective function than \cite{Elgeti2012}. Instead of full \gls{FEM} simulations, the authors employed analytical surrogate models obtained from previous numerical studies. Regarding the rheological behavior, they considered the non-isothermal flow of inelastic Bird-Carreau fluids. In a succeeding publication \cite{Rajkumar2018}, the authors proposed an updated version of the employed surrogate models, which incorporated the influence of the power law index as this proved to affect the system behavior the most. \\
Zhang \etal \cite{Zhang2019a} also leveraged \gls{NURBS} for parameterizing the inflow part of the extrusion die geometry and proposed two new objective functions based on geometrical considerations, which they compared to established ones from the literature. For the optimization, they constructed a surrogate model from a set of high-fidelity \gls{FEM} simulations with the help of the response surface methodology.

In this work, we propose to use \acrfull{RL} for optimizing the shape of flow channels inside profile extrusion dies. \gls{RL} is a method from the field of machine learning, in which an agent interacts with an environment, whilst trying to find a (optimal) strategy for solving a given problem (thereby mimicking the way how humans learn to accomplish a new task, namely by trial-and-error) \cite{Arulkumaran2017}. While not necessarily superior to conventional optimization algorithms for one single optimization problem, we expect \gls{RL} techniques to perform especially well when repeatedly applied to similar optimization tasks, since the agent learns a general strategy for solving a class of problems instead of just concentrating on the solution of a single problem \cite{Kaelbling1996}. This renders them promising for applications in the field of profile extrusion, where various dies often exhibit only slight differences in the preforming zone of the flow channel. 

At the latest after its breakthrough in playing Atari games \cite{Mnih2013}, \gls{RL} has become an established method for solving tasks where a valid solution strategy can be learned from repeated interactions, e.g., like playing computer games \cite{Vinyals2019} or tabletop games \cite{Silver2018}. Yet, the method has also gained attention in fields like robotics \cite{Kober2014}, control theory \cite{Busoniu2018}, and engineering design automation \cite{Dworschak2022}.

Recently, \gls{RL} has been applied to solve problems from the field of \gls{CFD}.
The review by Rabault \etal \cite{Rabault2020} highlighted the developments with respect to optimal design and control tasks, where they believe the empirical strategies found by \gls{RL} algorithms to be beneficial compared to traditional approaches. Two subsequent reviews by Garnier \etal \cite{Garnier2021} and its updated version by Viquerat \etal \cite{Viquerat2021Review} also investigated the impact of \gls{RL} onto the \gls{CFD} community, extending the application areas to not only numerical, but also experimental applications, considering works on heat transfer, drag reduction, microfluidics, and swimming. In the following, we focus on the advances of \gls{RL} for shape optimization, paying special attention to the employed learning algorithms as well as the geometry parameterization. \\
A first step has already been taken in 2008 by Lampton \etal \cite{Lampton2008}, who utilized a Q-learning algorithm for optimizing the shape of an airfoil with respect to lift, drag as well as momentum about the tip. The geometry of the airfoil was parameterized by four scalar quantities, which were modified by the \gls{RL} agent. In contrast to the approaches presented in the following, a Q-learning algorithm is restricted to a discrete action space, i.e., the set of admissible actions needs to be finite. This is because Q-learning algorithms estimate the value of each individual action for a given state and then derive a policy from these value estimates. In their review, Viquerat \etal \cite{Viquerat2021Review} titled approaches with discrete action spaces \textit{incremental shape optimization}. Different \gls{RL} algorithms have also been applied to perform incremental shape optimization in further publications: 
Yan \etal \cite{Yan2019} employed a \gls{DDPG} algorithm to optimize the placement and geometry of a missile fin with respect to the missile's lift-to-drag ratio under additional geometric and aerodynamic constraints. The geometry was updated directly, employing a mesh-update method based on radial-basis functions for running the \gls{CFD} simulations on the modified geometries.
The same \gls{RL} algorithm was leveraged in the work by Qin \etal \cite{Qin2021} to optimize compressor cascade blade profiles under multiple objectives, including a low total pressure loss, a large laminar flow area as well as constraints on the outlet flow angle. In order to obtain a flexible geometry parameterization, the authors perturbed an initial blade profile with Hicks-Henne bump functions.
The latter have also been used by Li \etal \cite{Li2021} to optimize the shape of supercritical airfoils under transonic flow conditions; however, only for additional local geometry modifications, while the overall airfoil shape has been described by a high-order Bernstein polynomial. The optimization itself was performed by a \gls{PPO} algorithm. \\
As an alternative to the incremental shape optimization approach, in \textit{direct shape optimization}, the design variables are not restricted to discrete values, meaning the agent can choose from a continuous action space. This approach has been used, e.g., by Viquerat \etal \cite{Viquerat2021Paper} to optimize two-dimensional shapes with respect to their lift-to-drag ratio. The shapes were parameterized by Bézier-curves and three different configurations with a different number of \glspl{DOF} (up to $12$ in total) were examined. The optimization was once again performed by a \gls{PPO} algorithm.
Concluding, we want to mention the work of Ghraieb \etal \cite{Ghraieb2022}, who recently introduced a framework for \gls{RL} in \gls{CFD}, using a Policy Based Optimization algorithm \cite{Viquerat2021PBO} to optimize both two and three-dimensional airfoil shapes with respect to their lift-to-drag ratio under moderately turbulent flow conditions. The airfoil shapes were represented once again by Bézier-curves and afterward immersed in a computational mesh, which was anisotropically adapted to align with the changing geometries.

In this paper, we present a thorough study of \gls{RL} applied to the problem of shape optimization of flow channels in profile extrusion dies. Therefore, we consider two different two-dimensional problems. First, we optimize a T-shaped geometry with respect to the mass-flow ratio between its two outflows, before we optimize a converging channel geometry with respect to a state-of-the-art flow-homogeneity criterion used in the literature for designing flow channels in profile extruders. The goal of this paper is to address the following three research questions:
\begin{enumerate}[label=(RQ\arabic*), ref=(RQ\arabic*)]
    \item \label{itm:RQ1}How can shape optimization of profile extrusion dies be realized by the means of \gls{RL} and are there obvious differences between these approaches?
    \item \label{itm:RQ2}How do the available learning algorithms differ with respect to the convergence and reproducibility of the training process?
    \item \label{itm:RQ3}How can the training time of the \gls{RL} agents be accelerated?
\end{enumerate}

After this introduction, the rest of the paper is structured as follows: \cref{sec:Methods} provides a brief introduction to the field of \gls{RL}. Since a general introduction is beyond the scope of this work, we will restrict ourselves to the methods relevant to this paper. Therefore, we will introduce the different employed learning algorithms, especially highlighting the differences, which might impact their learning behavior, before familiarizing the reader with the concept of learning in multiple environments. Next, in \cref{sec:Realisation}, we explain how we have applied the concept of \gls{RL} to our shape optimization problem. We present the framework used for generating the results, introducing the employed geometry parameterization, the governing equations of our model problem, as well as the interaction of the \gls{RL} agents with their environment. To answer our research questions, we investigate two test cases, which we present in more detail in \cref{sec:TJunction,sec:Channel}. The first test case (\cref{sec:TJunction}) considers a T-shaped geometry and optimizes it with respect to the mass-flow ratio between its outflows and can be seen as a proof-of-concept test case, as it allows to study the feasibility and correctness of the optimized geometries by visual inspection. The second test case (\cref{sec:Channel}) is more complex and aims towards a more realistic scenario by considering the flow homogeneity at the outlet of a converging channel geometry as the design criterion. Finally, our paper is concluded in \cref{sec:Conclusion} by summing up the most important findings with respect to the research questions posed in this introduction.
\section{Methods: Learning Algorithms and Vectorized Environment Training}
\label{sec:Methods}

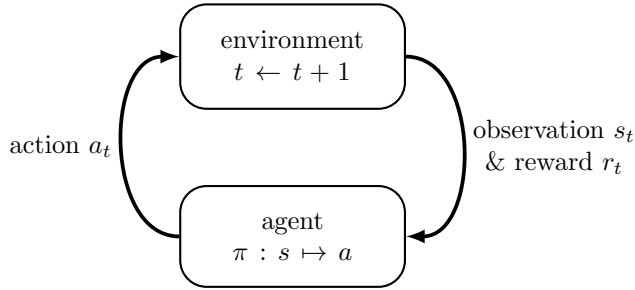
\begin{figure}[ht]
    \centering
    \begin{tikzpicture}
    \node[draw=black,thick,rounded corners=10pt,inner sep=10pt, align=center, text width=65, align=center] (env) at (0,0) {environment\\$t\leftarrow t+1$};
    \node[draw=black,thick,rounded corners=10pt,inner sep=10pt, below = of env, text width=65, align=center] (agent) {agent\\$\pi : s \mapsto a$};
    \draw[-latex,line width=0.5mm] (env.east) .. controls +(right:1) and +(right:1) .. (agent.east) node[midway, align=center, right] {observation $s_t$\\ \& reward $r_t$};
    \draw[-latex,line width=0.5mm] (agent.west) .. controls +(left:1) and +(left:1) .. (env.west) node[midway, align=center, left] {action $a_t$};
\end{tikzpicture}
    \caption{Interaction loop between an agent and an environment during training. In each interaction / training step $t$, the agent selects an action $a_t$ according to a policy $\pi$ based on observations of the current environment's state $s_t$ and a numerical reward signal $r_t$. This changes the state of the environment and generates the new observation and the new reward for the next step.}
    \label{fig:Methods:rl_loop}
\end{figure}

\gls{RL} is a sub-field of machine learning concerned with sequential decision-making. To solve such decision-making problems, an \textit{agent} interacts with an \textit{environment}, which encodes the problem to be solved. This interaction loop is visualized in \cref{fig:Methods:rl_loop}. The environment is in a distinct \textit{state} $s$ at any given time, which the agent can observe and, based on the observation, perform an \textit{action} $a$, to exert influence on the environment. This combination of observing and performing an action is typically called a \textit{step}. Solving the given problem involves taking a series of sequential steps, called an \textit{episode}, to achieve some predefined goal. The agent's behavior, i.e., the choice of which action to choose in a given state, is determined by a \textit{policy} $\pi: s \mapsto a$. 

This policy is learned during a training phase, where the agent interacts with the environment and receives \textit{rewards} $r$ depending on how well it is solving the task at hand. The collected experience consisting of states, actions, and rewards is used to iteratively improve the policy. During training, actions are selected either through the current best estimate of what a good policy would look like or through some exploration mechanism.

While \gls{RL} algorithms generally follow the framework described above to learn a policy, they differ substantially in their inner functioning and can be characterized based on a number of different properties. We explore these properties in the following \cref{subsec:Methods:Agents} and investigate which subset of algorithms is especially well-suited for shape optimization problems in profile extrusion in the remainder of this work. 

Since training \gls{RL} agents is dependent on large amounts of collected experience from the environment, the training process can be time-consuming. To address this issue, we further identify a suitable approach to reduce training times in \cref{subsec:Methods:MultiEnv}. 


\subsection{Agents - The \gls{RL} learning algorithms}
\label{subsec:Methods:Agents}

A major goal of our study is to investigate the suitability of different \gls{RL} algorithms to shape optimization problems in profile extrusion. Since different algorithms feature different characteristics, specific algorithms or subgroups of algorithms may emerge as especially suited in terms of training speed, stability, and final performance of the trained agents. 

\begin{figure}
    \centering
    \includegraphics[scale=0.65]{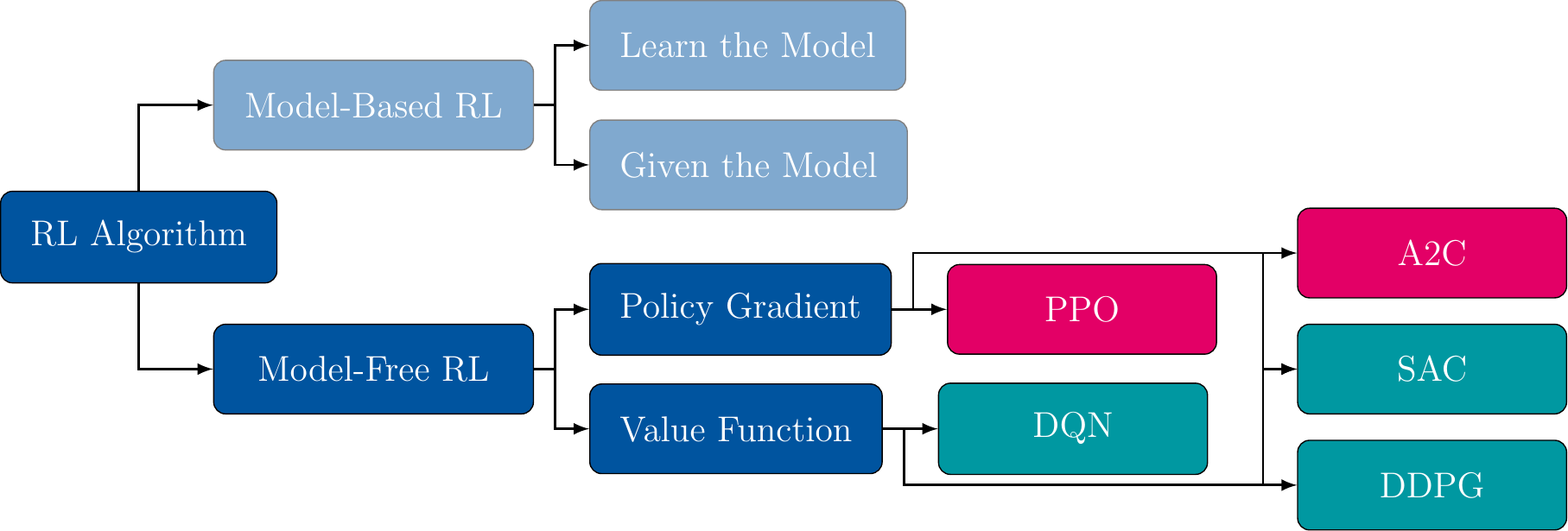}
    \caption[Agent Taxonomy]{Limited taxonomy of \gls{RL} agents. Blue items represent categories of agents. Turquoise items are agents trained with the off-policy method. Magenta items are agents trained with the on-policy method.}
    \label{fig:agent-taxonomy}
\end{figure}

In \cref{fig:agent-taxonomy}, a taxonomy of \gls{RL} agents is given, showing a range of different agents from diverse agent categories. Since this paper only studies model-free agents, no model-based examples are listed explicitly, but the category is depicted for completeness. Model-based \gls{RL} agents utilize a predictive model of the environment either during training or at decision time. This model of the environment can either be learned during training (Learn the Model) or given before starting the \gls{RL} training process (Given the Model). 

Model-free agents learn a strategy for solving their task without such an explicit predictive model. There are two main methods for learning this strategy. Policy gradient-based agents directly learn the policy, determining which actions to take in which states. In contrast, value function-based approaches learn a function predicting the expected return or reward of the possible actions. This value function is then utilized in the policy to decide which actions to perform. Examples for both approaches are given in the taxonomy, with \gls{PPO} \cite{Schulman2017} belonging to the class of gradient-based methods and \gls{DQN} \cite{Mnih2013} being a value function-based approach. The idea of policy gradient and value function-based methods can also be combined. These combined approaches commonly use an actor-critic structure, consisting of two parts. The critic learns a value function, which is then used to update the actor's policy function. Agents with this training approach are, e.g., \gls{A2C} \cite{Mnih2016}, \gls{SAC} \cite{Haarnoja2018}, and \gls{DDPG} \cite{Lillicrap2016}.
This combined approach alleviates some problems with pure policy gradient algorithms, which can have stability issues during training, but are generally faster to train than pure value function-based methods \cite{Konda2000}. \\
Agents can be additionally classified based on the data they use for training.
Here, one differentiates between \textit{on-policy} and \textit{off-policy} agents. 
On-policy algorithms can only use training samples generated with the agent's current policy and need to be discarded after each policy update. Consequently, on-policy methods can only use each training sample once, i.e., in the same training step in which it was generated.
In contrast, off-policy methods have no restrictions on the usability of training samples concerning the state of the agent's policy. They can reuse already collected samples even if they are generated with now outdated policy states. Thus, agents with this training method are more sample efficient \cite{Sutton2018}.


\subsection{Vectorized Environment Training - Accelerating training times}
\label{subsec:Methods:MultiEnv}

While \gls{RL} agents, once trained, can solve given problems very quickly, the training process itself tends to be time- and resource-intensive. This is especially problematic during the initial development phase, where multiple parameters are typically iteratively refined until a configuration is found that successfully solves the problem. The longer the time required to evaluate the effects of changes to the system, the slower the development progresses. Hence it is beneficial to explore approaches for reducing training times. 

One possibility to reduce training times is parallelization. As discussed in \cite{Rabault2019}, the speedup achievable by increasing the parallelization of the computations carried out within the environment (i.e., a \gls{CFD} simulation in their case) is typically limited. This is especially true for small calculations: Here, the startup costs and the communication overhead weigh more severe and are not necessarily dominated by the time required to perform the actual computations.
Instead, Rabault and Kuhnle \cite{Rabault2019} proposed a different parallelization approach, which does not provide the \gls{CFD} simulation with more computational resources, but creates multiple environments simultaneously, each running separate \gls{CFD} simulations. A single agent can then collect experiences from all of these environments in parallel and consequently collect a larger amount of experience in a fixed amount of time compared to the usual single environment setup. These environments are independent of each other, i.e., they do not communicate, so that communication only happens between the agent and each environment. In this configuration, the agent needs to be enabled to provide (different) actions to all environments in each training step and gather the training data consisting of observations, rewards, and actions. 
This method is known as \textit{vectorized environment training} in the field of \gls{RL}.

In our work, we follow the same strategy as outlined above to accelerate the training times of our agents. 
Our framework for \gls{RL}-based shape optimization -- which we introduce in more detail in the next section -- employs the open-source python package \texttt{stable-baselines3} \cite{Raffin2021} for the implementation of the \gls{RL} algorithms as well as realizing the vectorized environment training.

\section{Realisation - Shape Optimization via \gls{RL}}
\label{sec:Realisation}

\begin{figure}[ht]
    \centering
    \includegraphics[scale=0.85]{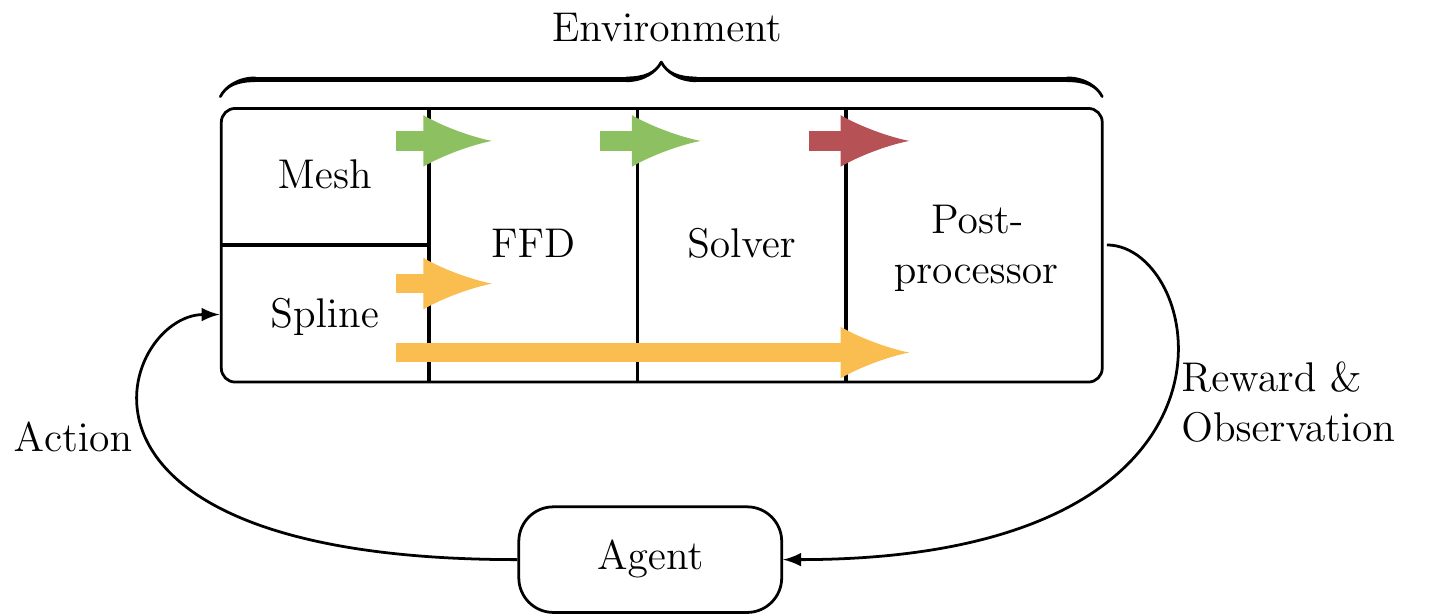}
    \caption{Visualization of the \gls{RL} interaction between an agent and the environment in our \texttt{releso} framework. The environment has been customized to our shape optimization problem and comprises the base mesh and the deformation spline used for the geometry parameterization, a \acrshort{FFD} module deforming the mesh, the solver computing the governing \gls{PDE} problem, and a component for postprocessing the simulation results to determine the reward and an observation of the \gls{CFD} environment for the agent. The arrows correspond to the information flows between the different components: Green arrows represent meshes, yellow arrows spline parameterizations, and red arrow stands for the simulation results. Based on the provided reward and observation, the agent chooses an action to modify the deformation spline of the \acrshort{FFD}.}
    \label{fig:Realisation:Releso}
\end{figure}

In this section, we cover the key aspects to leverage \gls{RL} for the shape optimization of profile extrusion die flow channels. Therefore, relevant \gls{RL} components, e.g., the environment as well as the agent's action and observation spaces, have to be defined accordingly. \cref{fig:Realisation:Releso} depicts the \gls{RL} interaction loop between the agent and our custom environment, which is defined in our framework \texttt{releso}, following the \gls{RL} environment interface provided by the state-of-the-art package \texttt{OpenAI Gym} \cite{Brockman2016}. For our application, this environment comprises the parameterization of the extrusion die flow channel's geometry and a \gls{CFD} simulation of the plastic melt inside the die.

The remainder of this section loosely follows the flow of information indicated by the arrows in \cref{fig:Realisation:Releso}: We begin with discussing different possibilities to define an agent's action space in \cref{subsec:Realisation:Actions}. This definition is closely related to the two \gls{RL}-based shape optimization approaches mentioned in the introduction. We explain how the agent can modify a geometry through its actions to enable \gls{RL}-based shape optimization for both cases. \cref{subsec:Realisation:FFD} then introduces the concept of \gls{FFD}, which we employ to parameterize our geometries. We conclude this section by stating the governing equations of plastic melt flow, which constrain our shape optimization problem, and comment on their numerical solution in \cref{subsec:Realisation:Environment}. 


\subsection{Actions: Shape optimization methods}
\label{subsec:Realisation:Actions}

As already mentioned in \cref{sec:Introduction}, two \gls{RL}-based shape optimization methods are described in \cite{Viquerat2021Review}, namely direct and incremental shape optimization. Which method can be used to optimize a geometry, depends on the definition of the agent's action space. Generally, \gls{RL} agents can either work with continuous or discrete action spaces, although there exist some algorithms that can work with both. In the case of continuous action spaces, the agent can choose action values from a continuous interval, while for discrete action spaces, it selects a value from a discrete set.

When applied to shape optimization problems, the agent needs to be able to modify the geometry with its actions. Its actions are thus defined such that they change the design variables, i.e., the \glspl{DOF} of the geometry parameterization. \\
Discrete actions are used in the incremental shape optimization approach. In each step of the \gls{RL}-loop, a single \gls{DOF} is incremented or decremented by a predefined amount inside the action's value interval. This provides the agent with two actions per \gls{DOF} in the problem definition.
In contrast, the direct shape optimization approach uses continuous actions. Here, each \gls{DOF} corresponds to a single action which is only constrained by the provided interval. While in the incremental approach only a single \gls{DOF} can either be incremented a decremented at a time, direct shape optimization allows changing all \glspl{DOF} at once in each step.

The choice of the action space consequently impacts the way how a shape is optimized: The incremental approach slowly moves toward the optimal geometry, modifying only a single design variable in each iteration, while the direct approach proposes the optimal geometry directly, i.e., ideally, in a single step. Therefore, we expect the training of agents for a direct optimization approach to be more challenging, both with respect to the number of training steps required and the design of the reward function. We will discuss the latter for each approach in the sections of the respective examples. 

Most agents introduced in \cref{subsec:Methods:Agents} do not support both shape optimization approaches described in this section. For the agent implementations of \texttt{stable-} \texttt{baselines3} used in this paper, \cref{tab:Realisation:AgentsShapeOptimizationMethodsCapabilities} provides an overview of their compatibility with the respective approach.

\begin{table}[ht]
    \centering
    \caption{Compatibility of the agents implemented in \texttt{stable-baselines3} with regard to the direct and incremental shape optimization method.}
    \begin{tabular}{lcc}
        \toprule
         Agent & Incremental & Direct \\
         \midrule
         \gls{PPO} & \checkmark & \checkmark \\
         \gls{DQN} & \checkmark & - \\
         \gls{SAC} & - & \checkmark \\
         \gls{A2C} & \checkmark & \checkmark \\
         \gls{DDPG} & - & \checkmark \\
         \bottomrule
    \end{tabular}
    \label{tab:Realisation:AgentsShapeOptimizationMethodsCapabilities}
\end{table}


\subsection{\gls{FFD}: Parameterizing the geometry}
\label{subsec:Realisation:FFD}

We have seen in the previous section how an agent can modify a given geometry parameterization through its actions. A crucial task in shape optimization however, is to first find a suitable parameterization. Since we want to employ the proposed \gls{RL}-based shape optimization method to a variety of extrusion-die flow-channel geometries in the future, the chosen parameterization should be generally applicable and not exploit any features that are restricted to a special geometry. Moreover, we want to control the resulting \glspl{DOF} of the parameterization, i.e., ideally it should allow to modify even intricate 3D geometries with relatively few parameters.

One possibility would be to directly modify the \gls{CAD} model of the flow channel geometry and generate a new mesh in every iteration. However, generating a mesh from a \gls{CAD} representation can be a challenging and computationally intensive task, which -- especially for realistic geometries -- cannot be easily automated \cite{Cottrell2009}. Normally, a mesh is only generated once for a geometry and thus the meshing costs also occur only once, which is not the case for shape optimization, where the geometry is gradually modified. Yet, the geometry only undergoes small changes between two succeeding design iterations, motivating the use of a less computationally expensive method.

\begin{figure}[ht]
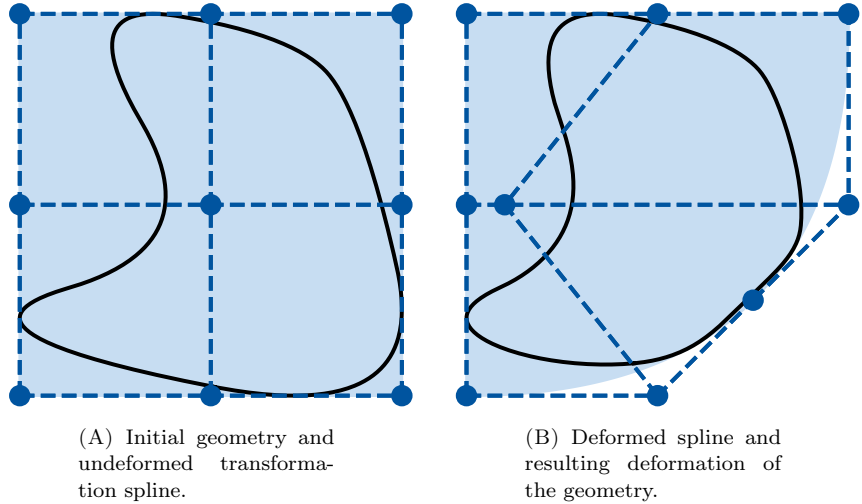

	\centering
	\subfloat[Initial geometry and undeformed transformation spline.]{
	    \label{fig:Realisation:FFD:Undeformed}
        \input{source/graphics/undeformed_fem_potato_w_spline.pgf}
	}
	\subfloat[Deformed spline and resulting deformation of the geometry.]{
	    \label{fig:Realisation:FFD:Deformed}
        \input{source/graphics/deformed_fem_potato_w_spline.pgf}
	}
	\caption{Visualization of the key idea of \gls{FFD}: An initial geometry (A) is transformed by modifying the control points (dark-blue dots) of a transformation spline (highlighted in light-blue) to obtain a deformed geometry (B).}
	\label{fig:Realisation:FFD}
\end{figure}

To this end, we propose the use of \gls{FFD} \cite{Sederberg1986}. Here, an initial geometry is meshed once and this base mesh is then deformed by a transformation function to represent geometric changes. This means that a mesh of the modified geometry can be generated without the need of first adapting the \gls{CAD} representation and then re-meshing it in each iteration. Splines (e.g., B-splines or \gls{NURBS} \cite{Piegl1995}) are a common choice for these transformation functions since they result in smooth deformations. Moreover, they allow the representation of deformations with a comparatively small amount of parameters. This is of great interest in both optimization and \gls{RL} as it helps to keep the dimensions of the design variables / action space low. To adapt the mesh, the position of each vertex in the mesh is transformed according to the transformation spline. This is visualized in \cref{fig:Realisation:FFD}. \cref{fig:Realisation:FFD:Undeformed} shows the initial geometry, embedded into the parameter space of the undeformed transformation spline depicted in light-blue. The control points of the transformation spline (the dark-blue colored dots) are the \glspl{DOF} of this geometry parameterization. Once they have been modified (e.g., through actions chosen by the \gls{RL} agent), the initial geometry can be deformed accordingly, as shown in \cref{fig:Realisation:FFD:Deformed}. To perform the \gls{FFD} within our environment, the open source python package \texttt{gustaf} \cite{gustaf} is used. 


\subsection{Solver: The \gls{PDE} problem governing the flow of molten plastic}
\label{subsec:Realisation:Environment}

In this work, we want to optimize flow channels inside profile extrusion dies with respect to different quantities of interest. Similar to classical shape optimization, the optimization objective needs to be re-evaluated after each geometry modification. This requires to solve the underlying equations that govern the flow of the highly-viscous plastic melt inside the extrusion die geometry $\Omega$. In our model, we assume a stationary, incompressible and isothermal flow. Since the Reynolds number characterising the flow in profile extrusion dies is typically very small, i.e., $Re \ll 1$, we neglect convection. Under these assumptions, the conservation laws of mass and momentum are given by the following \glspl{PDE}
\begin{subequations}
\label{eq:Realisation:StokesSystem}
\begin{align}
    \label{eq:Realisation:MassConservation} 
    \bm{\nabla} \cdot \bm{v} &= 0 \quad \text{in} \; \Omega \\ 
    \label{eq:Realisation:MomentumConservation} 
    -\bm{\nabla} \cdot \bm{\sigma} &= \bm{0} \quad \text{in} \; \Omega \eqdot 
\end{align}
\end{subequations}
Here, $\bm{v}$ denotes the unknown velocity vector and $\bm{\sigma}$ the Cauchy stress tensor. The latter needs to be related to the velocity vector via an additional equation to close the equation system \eqref{eq:Realisation:StokesSystem}. This is given by
\begin{subequations}
\label{eq:Realisation:ClosureEquations}
\begin{align}
    \label{eq:Realisation:CauchyStressTensor} 
    \bm{\sigma} &= 2\eta\bm{\varepsilon} -p\bm{I} \\
    \label{eq:Realisation:RateOfStrainTensor} 
    \bm{\varepsilon} &= \frac{1}{2}\left(\bm{\nabla}\bm{v} + \left(\bm{\nabla}\bm{v}\right)^T\right) \eqcolon 
\end{align}
\end{subequations}
where $p$ is the unknown pressure field. $\bm{\varepsilon}$ is referred to as the rate-of-strain tensor. In the case of Newtonian fluids, the viscosity $\eta$ is constant. However, this does not hold true for most plastic melts, where the viscosity depends on the shear rate. In our model, we want to include the shear-thinning behavior of molten plastic, i.e., the decrease of viscosity with increasing shear rates. Thus, we employ the semi-empirical relation found by Carreau \cite{Carreau1972}
\begin{subequations}
\label{eq:Realisation:MaterialLaws}
\begin{align}    
    \label{eq:Realisation:CarreauViscosity} 
    \eta &= \frac{A}{(1+B \dot{\gamma})^C} \eqcolon 
\end{align}
which relates the viscosity to the shear-rate 
\begin{align}
    \label{eq:Realisation:ShearRate} 
	\dot{\gamma} &= \sqrt{2\bm{\varepsilon}:\bm{\varepsilon}} \eqdot
\end{align}
\end{subequations} 
The parameters $A$, $B$, and $C$ are material-specific and assumed to be constant for our application. The values chosen in our test cases are given in \cref{tab:Realisation:MaterialProperties}.
\begin{table}[ht]
\centering
\caption[Material Properties]{Material properties of the shear-thinning material law for all test-cases.}
\label{tab:Realisation:MaterialProperties}
\begin{tabular}{lccc}
	\toprule
	Property & Symbol & Value & Unit \\
	\midrule
	zero-shear viscosity & $A$ & 10935 &  \si{\kg\per\meter\per\second} \\
	reciprocal transition rate  & $B$ & 0.433 &  \si{\per\second} \\
    slope of viscosity curve in pseudoplastic region & $C$ & 0.699 & - \\
	\bottomrule
\end{tabular}
\end{table}
To solve the \gls{PDE} system \eqref{eq:Realisation:StokesSystem}, proper boundary conditions need to be chosen. For our application, we impose no-slip boundary conditions on the walls of the profile extrusion die as well as a traction-free condition on the outflow:
\begin{subequations}
\label{eq:Realisation:BoundaryConditions}
\begin{align}
    \bm{v} &= \bm{0} && \text{on} \; \Gamma_\text{wall} \\
    \bm{\sigma} \cdot \bm{n} &= \bm{0} && \text{on} \; \Gamma_\text{out} \eqdot
\end{align}
\end{subequations}
The inflow boundary condition depends on the considered geometry and will thus be reported for each test case in \cref{sec:TJunction,sec:Channel}. The complete problem, consisting of the \glspl{PDE} \cref{eq:Realisation:StokesSystem}, the closure equations \cref{eq:Realisation:ClosureEquations}, the material law \cref{eq:Realisation:MaterialLaws}, and the boundary conditions \cref{eq:Realisation:BoundaryConditions} is solved using an in-house simulation code. Therein, the continuous \gls{PDE} system is discretized using a stabilized \gls{FEM} approach \cite{Tezduyar1992a} with linear basis functions for the unknown pressure and velocity fields.  

The computation of a solution to the model equations is embedded into our custom environment. In every \gls{RL} step, i.e., after each geometry modification, our in-house simulation code is invoked to calculate new solutions for the velocity and pressure fields. Based on these, the numerical design criterion can be evaluated.


So far, we have only introduced the components of our \gls{RL} framework, which are test case-independent. However, some components like, e.g., the reward function or the agent's observation space depend on considered problem. Thus, they are defined in the following \cref{sec:TJunction,sec:Channel}, where two selected application cases are presented, for which we want to answer research questions \cref{itm:RQ1,itm:RQ2,itm:RQ3} as posed in the introduction.
\section{An introductory example: Optimizing a T-shaped geometry with respect to the mass-flow ratio between its outflows}
\label{sec:TJunction}

As a first test case, we consider a simple two-dimensional T-shaped geometry, serving as an introductory example to assess the capabilities of the proposed \gls{RL}-based shape optimization framework. Already introduced in a prior work by the authors \cite{Wolff2022} to investigate the applicability and reproducibility of the incremental shape optimization approach learned by a \gls{PPO} agent, we extend this application here to a direct shape optimization approach and compare the performance of different agents. 

\begin{figure}
    \centering
    \input{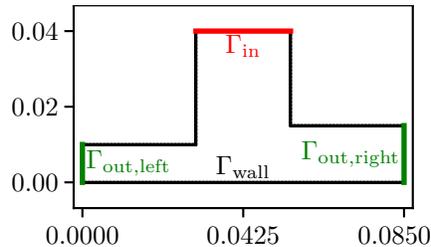}
    \caption{Geometry and boundaries of the T-shaped geometry.}
    \label{fig:TJunction:Geometry}
\end{figure}

The geometry to be optimized is depicted in \cref{fig:TJunction:Geometry}. The general learning task is stated as follows: Optimize the geometry $\Omega$ such that a desired mass-flow ratio $\mu^\star$ between the mass flows through the left and right outflow boundaries is achieved. While this simple objective does not resemble the overall goal to optimize an extrusion die geometry with respect to the flow homogeneity at the outflow, it allows us to intuitively assess the correctness of the optimized geometries. 

To make the learning process more general, we do not train the agents to optimize the geometry for a single target mass-flow ratio, but instead, randomly select a new one from a predefined interval according to a uniform distribution, i.e., $\mu^\star \sim \mathcal{U}(0.1,2)$, at the beginning of each episode. Thus, after training an agent, it should be able to generate optimal shapes for a variety of mass-flow ratios.

In the rest of this section, we will first introduce the parameterization of the test case's geometry, i.e., the agent's action space. We continue with a description of how the optimization objective can be determined from the results of the \gls{CFD} simulation. This is followed by a section on reward shaping, which is crucial as the reward signal guides the agent toward learning a good strategy. 
For the results, we train different agents with both an incremental and a direct strategy and compare the learning processes with respect to their convergence behavior. We will furthermore compare the general behavior of both approaches. To wrap this section up, we investigate the vectorized environment training with regard to a potential decrease in the real-world training time (wall time) of the agent.


\subsection{Geometry parameterization}
\label{subsec:TJunction:GeometryParameterization}

The T-shaped geometry depicted in \cref{fig:TJunction:Geometry} represents a flow separator with one inflow and two outflows. The boundary conditions on the wall $\Gamma_{\text{wall}}$ and at the outflow  $\Gamma_{\text{out}}=\Gamma_{\text{out},\text{left}} \cup \Gamma_{\text{out},\text{right}}$ are given by \cref{eq:Realisation:BoundaryConditions}. The boundary condition on $\Gamma_{\text{in}}$ is given by
\begin{align}
    \label{eq:TJunction:InflowBC}
    \bm{v} &= \begin{pmatrix}0\\-0.45\end{pmatrix} && \text{on} \; \Gamma_\text{in} \eqdot
\end{align}

\begin{figure}
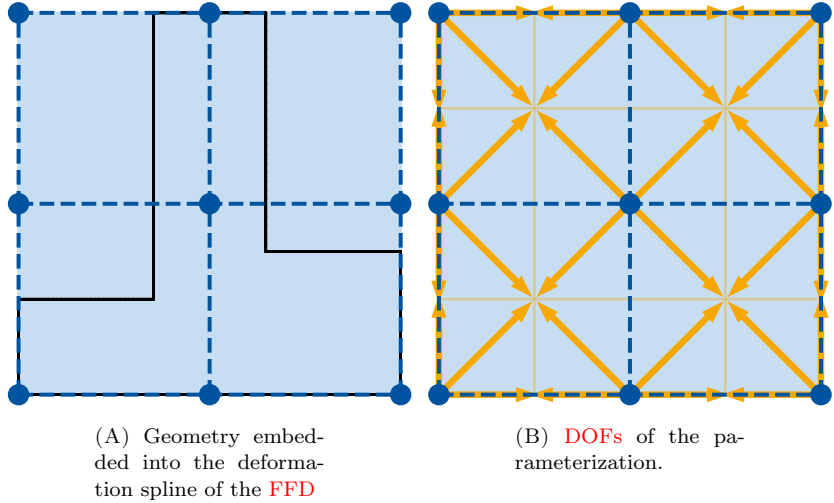

    \centering
    \subfloat[Geometry embedded into the deformation spline of the \gls{FFD}]{\input{source/graphics/tjunction_w_spline.pgf}}
    \subfloat[\glspl{DOF} of the parameterization.]{\input{source/graphics/tjunction_w_spline_movement.pgf}}
    \caption{(A) shows the deformation spline used for the parameterization of the T-shaped geometry. To make the \gls{FFD} more generic, the actual geometry is scaled to the parametric space of the transformation spline before applying geometric modifications. Additionally, the possible movement of the control points is illustrated by the orange arrows in (B).}
    \label{fig:TJunction:Spline}
\end{figure}

The deformation spline is a two-dimensional B-spline with second order basis functions in both parametric dimensions. It features nine control points in total, arranged as shown in \cref{fig:TJunction:Spline}. This gives the agent $18$ \glspl{DOF}, two for each control point, as the control points can move in $x$ and $y$ direction. Agents following an incremental shape optimization approach can choose from $36$ discrete actions while agents utilizing the direct optimization approach operate with an $18$-dimensional action space.

The base mesh for the \gls{FFD} consists of $4759$ triangular elements and $2515$ nodes.


\subsection{Observations}
\label{subsec:TJunction:Observations}

Since we want to optimize the T-shaped geometry with respect to desired mass-flow ratio $\mu^\star$, the agent needs to be provided with information about the mass-flow ratio for a given geometry. Therefore, we can compute the mass-flow ratio from the results of the \gls{CFD} simulation. The mass-flow ratio between the left and right outflow at a certain step $t$ is defined as
\begin{align}
    \label{eq:TJunction:MassFlowRatio}
    \mu_t &= \frac{(\dot{m}_\mathrm{out,left})_t}{(\dot{m}_\mathrm{out,right})_t} \eqcolon
\end{align}
where the mass flows over each outflow boundary $\Gamma_\mathrm{out,left}$ and $\Gamma_\mathrm{out,right}$ can be computed by integrating the velocity field $\bm{v}$ over each boundary
\begin{align}
    \label{eq:TJunction:MassFlow}
    \dot{m}_{(\cdot)} &= \rho \int_{\Gamma_{(\cdot)}} \bm{v} \cdot \bm{n} \; d S \eqdot
\end{align}
Here, $\rho$ denotes the density of the molten plastic, which is constant due to the incompressibility assumption. However, as long as we are only considering mass-flow ratios according to \cref{eq:TJunction:MassFlowRatio}, the density is irrelevant since it cancels out.

In each step $t$, the agent is now provided with the mass-flow ratio $\mu_t$, the target mass-flow ratio $\mu^\star$, and the control points of the transformation spline.


\subsection{Reward shaping}
\label{subsec:TJunction:RewardShaping}

Reward shaping is one of the most crucial parts in \gls{RL}, as the reward is responsible for guiding the agent toward a good strategy for solving the problem. However, this is not a straightforward task, as it greatly depends on the problem under consideration as well as the shape optimization strategy utilized. In the following sections, we present reward functions for both shape optimization strategies.


\subsubsection{Direct Approach}
\label{subsec:TJunction:RewardShaping:Direct}

For the direct approach, we first define the deviation of the current step's mass-flow ratio from the desired target ratio according to
\begin{align}
\label{eq:TJunction:MassFlowError}
	e_{(\cdot)} &= \lvert \mu_{(\cdot)} - \mu^\star \rvert \eqdot
\end{align}
This quantity is then used with a Boolean value indicating the success of the simulation to shape the following reward function
\begin{align}
\label{eq:TJunction:Reward:Direct}
	r_t &= \begin{cases}
	    -10 &  \text{simulation failed} \\
		-e_t & e_t \geq \epsilon \\
		5 & e_t < \epsilon
	\end{cases}
	\eqdot
\end{align} 
This reward function is non-sparse, i.e., the agent receives a reward signal in every step and not only after it accomplishes its task. A failed simulation (which should only occur in the case of a tangled mesh) is severely penalized by a fixed value of $-10$. In contrast, the agent receives a constant reward of $5$ if the deviation from the goal ratio falls below the given acceptance threshold $\epsilon$. Since we ideally want the agent to reach the goal in a single step in the direct approach, a negative reward is given for every step in which the agent does not reach the goal. To provide the agent with some information on how close the current shape is to the optimal shape, this negative reward corresponds to the deviation from the goal ratio instead of, e.g., a constant negative reward. 


\subsubsection{Incremental Approach}
\label{subsec:TJunction:RewardShaping:Incremental}

For this test case, the incremental approach has already been investigated in \cite{Wolff2022} and we employ the same reward function here. It depends on the current target mass-flow ratio $\mu^\star$, the mass-flow ratios $\mu_t$ and $\mu_{t-1}$, as well as the deviations from the target mass-flow ratio $e_t$ and $e_{t-1}$ at steps $t$ and $t-1$
\begin{align}
\label{eq:TJunction:Reward:Incremental}
	r_t &= \begin{cases}
		-10 & \text{simulation failed} \\
		-0.5 & e_t > e_{t-1} \\
		-0.2 & e_t = e_{t-1} \\
		\frac{\lvert \mu_t - \mu_{t-1} \rvert}{e_{t-1}} & e_t < e_{t-1} \wedge e_t \geq \epsilon \\
		5 & e_t < \epsilon \\
	\end{cases}
	\eqcolon
\end{align} 
Similar to the reward function in the direct approach \cref{eq:TJunction:Reward:Direct}, a reward of $-10$ is only generated, if the episode is considered unsuccessful and terminated prematurely. Likewise, a reward of $5$ corresponds to a successful solution of the task. With the incremental approach, the agent is expected to optimize the geometry in multiple subsequent steps. Here, we can distinguish between three possible scenarios: If a modification of the geometry results in a bigger deviation from the desired target, i.e., $e_t > e_{t-1}$, the agent is penalized with a negative reward of $-0.5$. If the agent did not make progress, i.e., $e_t=e_{t-1}$, it also receives a penalty, amounting to $-0.2$. In case of improvement concerning the desired target mass-flow ratio, the agent is rewarded proportional to the relative improvement compared to the previous step. This asymmetric definition of the reward has been chosen to prevent the agent from undertaking sequences of actions that would hinder it in increasing the accumulated reward. Instead, it obtains direct feedback that provides knowledge about the quality of the previous action regarding the desired goal.


\subsection{Agent Comparison}
\label{subsec:TJunction:AgentComparison}

In this section, we compare different learning algorithms following both direct and incremental optimization strategies to investigate their suitability and behavior with respect to training effort and convergence. Within the whole section, each experiment is performed twice to show consistency among repeated training runs.


\subsubsection{Direct Approach}
\label{subsec:TJunction:AgentComparison:Direct}

As shown in \cref{tab:Realisation:AgentsShapeOptimizationMethodsCapabilities}, four algorithms have the capability to be trained with a direct optimization approach, namely, \gls{PPO}, \gls{A2C}, \gls{SAC}, and \gls{DDPG}. All training runs shown here were conducted with the same parameterization of the environment using the default parameters of the algorithms according to the \texttt{stable-baselines3} package. The results for each agent are shown for up to \SI{150}{\kilo{}} training steps\footnote{The experiments were conducted for more steps, but all relevant information is contained in this interval. The complete dataset can be provided by the authors upon request.}. 

\begin{figure}[ht]
    \centering
    \input{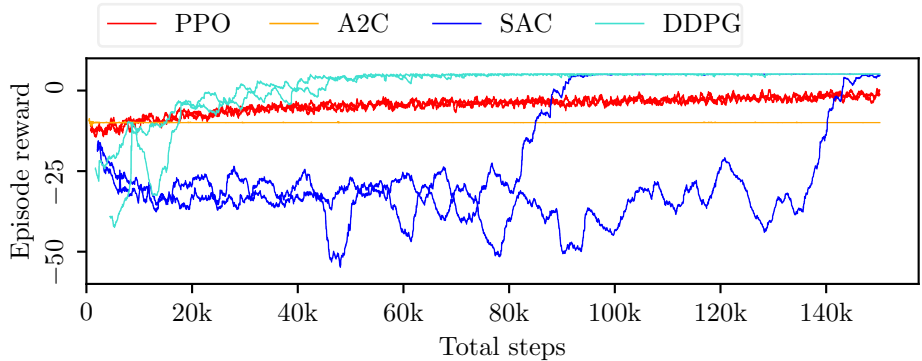}
    \caption{Comparison of different algorithms trained to optimize the T-shaped geometry following a direct strategy with respect to the episode reward over the trained steps. Each run was repeated twice as indicated using the same color.}
    \label{fig:TJunction:AgentComparison:Direct:Steps}
\end{figure}

In \cref{fig:TJunction:AgentComparison:Direct:Steps}, the progression of the agents' episode reward over the trained steps is shown.
The algorithms differ significantly with respect to the convergence of the episode reward. First, it can be noticed that \gls{DDPG} and \gls{PPO} manage to steadily increase their reward, albeit on different scales. Especially in the early training phase ($t\leq \SI{20}{\kilo{}}$), the \gls{DDPG} agents show more volatility compared to the \gls{PPO} agents. However, the \gls{DDPG} agents converge faster to the optimal reward value of $5$ (see \cref{eq:TJunction:Reward:Direct}). The reward curves corresponding to the two \gls{SAC} agents exhibit three distinct phases: Initially, the rewards stay more or less constant for roughly $\SI{40}{\kilo{}}$ steps. Afterward, the rewards start to oscillate, culminating in sharp jumps that increase the episode rewards to their optimal value of $5$. After the jumps, the episode rewards even out at this value. Noteworthy is that the sharp jumps occur after a totally different amount of training steps. This suggests that the \gls{SAC} agent is more susceptible to random variations during the training. The training runs conducted with the \gls{A2C} agent show no learning progression at all. While at first sight, this seems to suggest an inability of this algorithm to learn a valid strategy for this problem, it may also stem from the fact that the reward function shaped in \cref{eq:TJunction:Reward:Direct} does not work well for this particular algorithm. \\
In terms of consistency, of all agents that successfully learned a strategy on this test case, the \gls{PPO} agents show the smallest variation among repeated runs. In contrast, the biggest variations can be observed for the \gls{SAC} agents, as already discussed above. 

To answer \cref{itm:RQ2} properly, we also need to compare the different algorithms with respect to their total training time. The wall-clock times are reported in \cref{tab:TJunction:AgentComparison:Direct:WallTimes}.
\begin{wraptable}{r}{5cm} 
    \centering
    \caption{Wall-clock times of the agents trained with the direct optimization method to optimize the T-shaped geometry.}
    \begin{tabular}{lcc}
        \toprule
         Agent & Max. training time \\
         \midrule
         \gls{PPO} & 26.1 \si{\hour} \\
         \gls{A2C} & 46.5 \si{\hour} \\
         \gls{SAC} & 26.0 \si{\hour} \\
         \gls{DDPG} & 33.5 \si{\hour} \\
         \bottomrule
    \end{tabular}
    \label{tab:TJunction:AgentComparison:Direct:WallTimes}
\end{wraptable}
One can directly spot differences in the efficiency, as the different algorithms require different amounts of time to complete the demanded \SI{150}{\kilo{}} steps. As before, \gls{A2C} performs worst when measured in this metric. Of all algorithms exhibiting convergent behavior with respect to the episode reward, the \gls{DDPG} agent takes the longest to perform the \SI{150}{\kilo{}} steps. However, it would be possible to stop the training much earlier since it is already converged after roughly \SI{45}{\kilo{}} training steps as seen in \cref{fig:TJunction:AgentComparison:Direct:Steps}. The wall-clock times of \gls{SAC} and \gls{PPO} are equally long. However, the inconsistent behavior evident in \cref{fig:TJunction:AgentComparison:Direct:Steps} could also be observed with respect to the training times of the two runs. Both \gls{PPO} runs consistently took roughly \SI{26}{\hour}, but it needs to be mentioned that this algorithm did not fully converge, i.e., the optimal reward was not yet reached within the fixed amount of \SI{150}{\kilo{}} training steps. 

This investigation shows that it is important to not only focus on the training time alone, but also take the reward into account: In terms of training time, both \gls{PPO} runs outperform the \gls{DDPG} agents. However, the \gls{DDPG} agents accumulate their reward faster, thus amortizing the increased duration of the individual training steps to some extent.


\subsubsection{Incremental Approach}
\label{subsec:TJunction:AgentComparison:Incremental}

For the incremental optimization strategy, the employed \gls{RL} package \texttt{stable-baselines3} provides three algorithms that can operate on discrete action spaces (see \cref{tab:Realisation:AgentsShapeOptimizationMethodsCapabilities}), namely \gls{PPO}, \gls{A2C}, and \gls{DQN}. These will be compared in the following.

\begin{figure}[ht]
    \centering
    \input{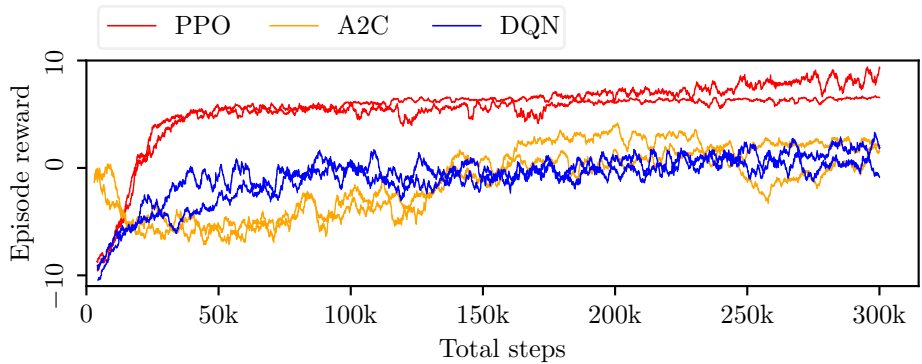}
    \caption{Comparison of different algorithms trained to optimize the T-shaped geometry following an incremental strategy with respect to the episode reward over the trained steps. Each run was repeated twice as indicated using the same color.}
    \label{fig:TJunction:AgentComparison:Incremental:Steps}
\end{figure}

\cref{fig:TJunction:AgentComparison:Incremental:Steps} again depicts the episode reward over the trained steps, for up to \SI{300}{\kilo{}} steps\footnote{The experiments were conducted for more steps, but all relevant information is contained in this interval. The complete dataset can be provided by the authors upon request.}. The \gls{PPO} agent exhibits a steep learning curve, especially during the early training phase ($t\leq \SI{50}{\kilo{}}$) and quickly converges to the optimal reward of $5$. Afterward, the curve of the episode reward flattens and oscillates around this value. Both \gls{PPO} runs show very similar behavior. The other two algorithms do not converge in the allotted time; nevertheless their episode reward curves slowly increase over the depicted training steps. It is worth noting that here, the \gls{A2C} agent manages to learn a strategy for increasing its episode reward -- in contrast to its application with the direct optimization strategy. 
A possible reason behind this behavior as well as an in-depth comparison of the different agent types will be shortly discussed in \cref{subsec:TJunction:AgentComparison:LearningBaheviorA2C}. 
One interesting anomaly can be observed from \SI{250}{\kilo{}} steps onward: One of the \gls{PPO} training runs diverges and finds a way to increase its episode reward again. 
This can be explained with a non-optimal reward function as the agent evidently finds a way to exploit \cref{eq:TJunction:Reward:Incremental} to increase the reward beyond the optimal value of $5$. Exploitation of that sort can be prevented in two possible ways: Either the reward function needs to be improved or the training of the agent needs to be stopped as soon as it is converged.

\begin{wraptable}{r}{5cm} 
    \centering
    \caption{Wall-clock times of the agents trained with the incremental optimization method for the T-junction use case.}
    \begin{tabular}{lcc}
        \toprule
         Agent & Max. training time \\
         \midrule
         \gls{PPO} & 49.9 \si{\hour} \\
         \gls{A2C} &  45.1 \si{\hour}\\
         \gls{DQN} & 45 \si{\hour} \\
         \bottomrule
    \end{tabular}
    \label{tab:TJunction:AgentComparison:Incremental:WallTimes}
\end{wraptable}

In contrast to the results with the direct optimization strategy, the wall time comparisons for the incremental approach (see \cref{tab:TJunction:AgentComparison:Incremental:WallTimes}) reveal less pronounced differences between the different algorithms. The only outlier is one of the \gls{PPO} agent runs, which takes about \SI{5}{\hour} longer than the other. This can be attributed to the additional computational overhead associated with the time required to reset an episode: The \gls{PPO} training run, which does not diverge, needs to be reset more often, since it needs fewer steps per episode to accomplish its task than its counterpart, which just tries to exploit the reward function to increase its reward even further.

Examples of optimized geometries obtained from a \gls{PPO} agent following an incremental shape optimization approach are included in \cref{app:Appendix:OptimizedGeometry:TJunction}.

\subsubsection{A remark on the learning behavior of \gls{A2C}}
\label{subsec:TJunction:AgentComparison:LearningBaheviorA2C}

Although the \gls{A2C} agent did not manage to learn a good policy for optimizing the T-shaped geometry with a direct strategy, it did when following an incremental approach. Even more surprisingly -- as we will show in \cref{subsec:Channel:AgentComparison:Direct} -- the \gls{A2C} agent also managed to learn a strategy for optimizing the geometry of the second test case with the direct shape optimization method. 
While performing further experiments not contained in this paper, it was found that the \gls{A2C} agent manages to find and exploit a local optimum in the reward function \cref{eq:TJunction:Reward:Direct}. This local optimum always provides a constant penalty of $-10$ if the agent generates a tangled mesh, no matter in which training step. So if the agent generates a tangled mesh as fast as possible, i.e., in the first step, without accumulating additional negative rewards before, the episode reward converges to a value of $-10$. After this discovery, we adapted the reward function to resolve this issue. In an additional experiment with the adapted reward function, an \gls{A2C} agent following a direct shape optimization approach was also successfully trained for the T-shaped geometry. The results are not published in this paper since the change in the reward function would have required reevaluating all other experiments to make the results comparable. Nevertheless, we considered this finding worth mentioning here. 


\subsection{Vectorized Environment Training}
\label{subsec:TJunction:MultiEnv}

As we have seen in the previous sections, training an \gls{RL} agent can require a significant amount of time, depending on the employed optimization strategy as well as the learning algorithm. In this section, we want to address \cref{itm:RQ3} and see if the wall training time can be reduced by letting the agent interact with multiple environments. We will investigate this research question here exemplarily for the incremental optimization approach. \\
The previous results showed that the \gls{PPO} agent trained with an incremental optimization strategy has a stable and reproducible learning curve if the training is stopped after the algorithm is converged. Therefore, this combination was chosen to investigate the vectorized environment training. Even though only results for the \gls{PPO} agent are shown here, the other algorithms were also tested and behaved similarly. The conducted experiments include a comparison of training runs using one, two, four, and eight vectorized environments. Each agent is trained for a maximum number of \SI{100}{\kilo{}} steps.

\begin{figure}[ht]
    \centering
    \input{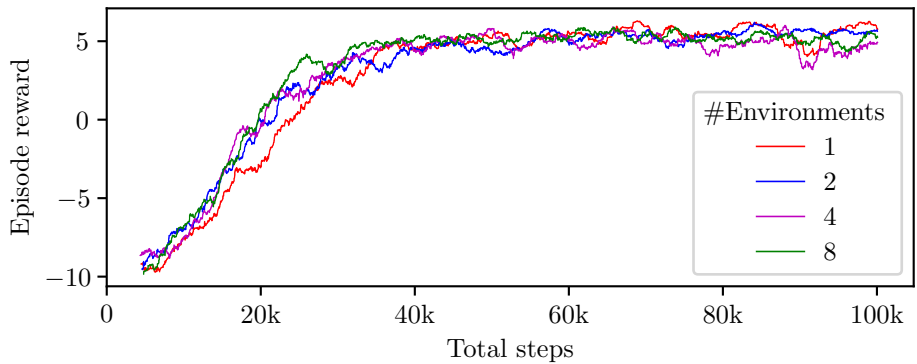}
    \caption{Episode reward over training steps of a \gls{PPO} agent following an incremental strategy for optimizing the T-shaped geometry when interacting with different numbers of environments.}
    \label{fig:TJunction:MultiEnv:Steps}
\end{figure}

\cref{fig:TJunction:MultiEnv:Steps} visualizes the agent's episode reward over the trained steps. It shows, that the training behavior does not significantly change if the number of environments is increased.

\begin{figure}[ht]
    \centering
    \input{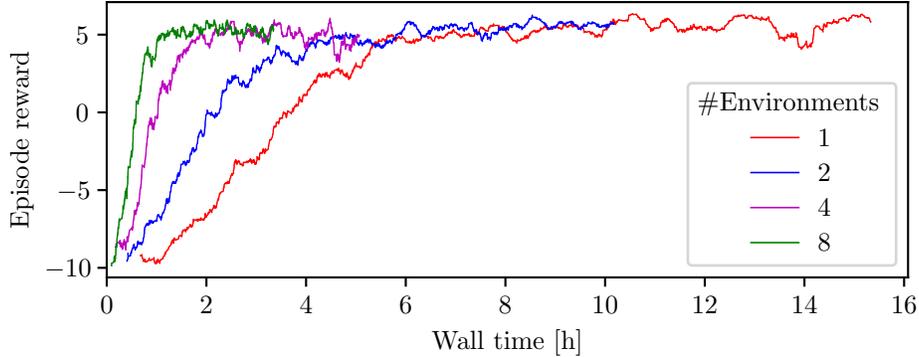}
    \caption{Episode reward over wall time of a \gls{PPO} agent following an incremental strategy for optimizing the T-shaped geometry when interacting with different numbers of environments.}
    \label{fig:TJunction:MultiEnv:WallTime}
\end{figure}

This however changes completely, if the episode reward is plotted over the wall time instead, as depicted in \cref{fig:TJunction:MultiEnv:WallTime}. Here, a significant decrease in the total training time can be observed. While not scaling perfectly, i.e., halving the training time with every doubling of the amount of environments, an overall reduction from about \SI{15.33}{\hour} with one environment to \SI{3.33}{\hour} with eight environments can be observed. This corresponds to a wall time reduction of about \SI{78}{\percent}, which answers \cref{itm:RQ3} for this application case.

The next section introduces a more realistic example with respect to the optimization of flow channels in profile extrusion dies.
\section{Towards a more practical application: Optimizing a converging channel with respect to the flow homogeneity}
\label{sec:Channel}

\begin{figure}[ht]
    \centering
    \input{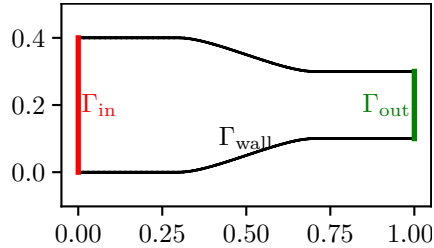}
    \caption{Geometry and boundaries of the converging channel geometry.}
    \label{fig:Channel:Geometry}
\end{figure}

In this test case, an agent is trained to optimize a two-dimensional converging channel geometry with respect to the homogeneity of the flow at the outlet. The initial geometry is depicted in \cref{fig:Channel:Geometry}. In real profile extrusion lines, this quantity has the largest influence on the quality of the manufactured part \cite{Hopmann2016}. Since we want to apply the \gls{RL}-based shape optimization method to realistic flow channel geometries in future research, with this test case, we want to investigate the feasibility of this method when employed with a state-of-the-art design criterion.
Compared to the previous one, this test case is especially challenging for two reasons, i.e., the generalization to a whole class of geometries and the inability to reach the optimum. \\
Generalizing the task of optimizing the flow homogeneity would require creating many diverse geometries. Otherwise, the agent could not learn a strategy applicable to all. If only a single geometry is presented during training, the agent only learns to solve this single problem. However, this would defeat the main attraction of this method, namely to train an agent for a more generalized task and then be able to solve specific tasks inside the generalized task's envelope faster than conventional optimization methods. \\
The inability to reach the optimum stems from the fact that the velocity profile with optimal flow homogeneity would be a block profile. However, such a velocity profile cannot be achieved in real profile extrusion lines since the plastic melt adheres to the channel walls. 
This, in turn, presents a problem for our \gls{RL} approach: Our experiments showed that the agent learns best if it tries to reach a discrete goal within each episode and if the learning is terminated upon achievement. If the episode is not ended (i.e., when the optimum cannot be reached), the agent will continue performing actions. At least with the incremental strategy, these additional actions will eventually move the state away from an already good solution, incurring negative rewards. One possible remedy would be to switch to sparse rewards, i.e., only generate a reward if the state reached the optimal flow homogeneity. In order to end the episode at or close to the optimal geometry, the best flow homogeneity achievable needs to be determined with the used geometry parameterization, creating a cyclic dependency. To overcome this issue, this paper used a preliminary training run to empirically determine the best achievable value of the quality criterion. Of course, this does not guarantee the optimality of this value, but possibly a sufficiently good approximation since more than \SI{100}{\kilo{}} designs have been evaluated. Following this approach, we found that the numerical value of the optimal homogeneity quality criterion $q^{\star} = 0.358$ is very close to the quality criterion of the initial channel geometry.

Therefore, the learning task for this test case can be stated as: Optimize the geometry $\Omega$ such that the flow homogeneity at the outflow boundary amounts to $q^{\star}$. To increase the generalizability of this test case, the initial geometry depicted in \cref{fig:Channel:Geometry} is deformed at the beginning of each episode by a randomly chosen parameterization of the same spline which is used by the agent. By doing this, the agent solves the previously stated learning task on a different geometry in each episode, thus learning a more general strategy. At the same time, this also circumvents the issue that the original geometry already provides a quantitatively good homogeneity, as far as our numerical experiments were able to show. 

The rest of this section largely mimics the structure of \cref{sec:TJunction}. For the results, however, an additional section is included, where we compare the direct and incremental shape optimization approach for this test case.


\subsection{Geometry parameterization}
\label{subsec:Channel:GeometryParameterization}

The converging channel geometry is depicted in \cref{fig:Channel:Geometry}. The boundary conditions on $\Gamma_\mathrm{wall}$ and $\Gamma_\mathrm{out}$ are once again chosen according to \cref{eq:Realisation:BoundaryConditions} and we also employ a block profile on $\Gamma_\mathrm{inflow}$ similar to \cref{sec:TJunction}, except that we rotate it by \SI{90}{\degree} to fit the changed orientation of the inflow boundary
\begin{align}
    \label{eq:Channel:InflowBC}
    \bm{v} &= \begin{pmatrix}0.45\\0\end{pmatrix} && \text{on} \; \Gamma_\text{in} \eqdot
\end{align}

\begin{figure}[ht]
    \centering
    \input{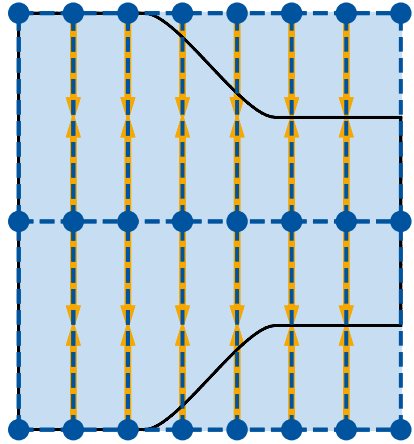}
    \caption{Deformation spline used for the parameterization of the channel geometry. Additionally, the possible movement of the control points is illustrated by the orange arrows.}
    \label{fig:Channel:Spline}
\end{figure}

The geometry is parameterized by a two-dimensional B-spline with second-order basis functions in both parametric directions. Its $24$ control points are arranged as shown in \cref{fig:Channel:Spline}. Here, we restrict the movement of the control points and only allow changes perpendicular to the main flow direction. Moreover, we fix the position of those control points, which would modify the position of the inflow or outflow. In total, we end up with $18$ control point coordinates that can move in $y$-direction. Agents following an incremental shape optimization approach thus choose from a set of $36$ discrete actions, while agents pursuing a direct optimization approach have an $18$-dimensional action space available.

The base mesh for the \gls{FFD} consists of $3886$ triangular elements and $2040$ nodes.


\subsection{Observations}
\label{subsec:Channel:Observations}

The optimization objective is to achieve a flow that is as homogeneous as possible at the outflow.
\begin{figure}[ht]
    \centering
    \input{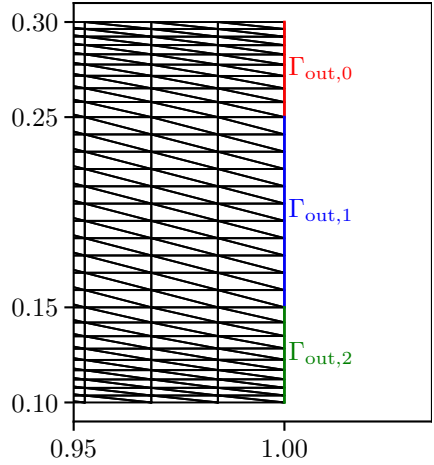}
    \caption[Channel Outflow Patches]{Outflow boundary of the converging channel geometry. The outflow is divided into three patches $\Gamma_{\text{out}, i}$, for which the flow-homogeneity criterion is evaluated.}
    \label{fig:Channel:OutflowPatches}
\end{figure}
Different methods to quantify the flow homogeneity have been studied in the literature \cite{Elgeti2012,Rajkumar2018}. They all have in common that they divide the outflow boundary into non-overlapping so-called \textit{patches} (highlighted in different colors in \cref{fig:Channel:OutflowPatches}) and 
evaluate a quality criterion on each patch. This patch-wise quality criterion usually relates the average velocity on these patches 
\begin{equation}
    \label{eq:Channel:VelocityOnPatch}
	v_{\text{out},i} = \frac{\dot m_{\text{out},i}}{\rho A_{\text{out},i}}
\end{equation}  
to the average velocity across the whole outflow boundary 
\begin{equation}
    \label{eq:Channel:AverageVelocity}
	v_\text{avg} = \frac{\dot m_\text{out}}{\rho A_\text{out}} \eqdot
\end{equation} 
Therein, the mass flows $\dot m_{(\cdot)}$ are computed according to \cref{eq:TJunction:MassFlow}, while $A_{(\cdot)}$ denotes the area of the respective boundary segment $\Gamma_{(\cdot)}$. The flow homogeneity is then quantified by the means of the patch-wise ratio between these velocities as
\begin{equation}
    \label{eq:Channel:VelocityRatio}
	\omega_{\text{out},i} = \frac{v_{\text{out},i}}{v_\text{avg}} \eqdot
\end{equation} 
Building on \cref{eq:Channel:VelocityOnPatch,eq:Channel:AverageVelocity,eq:Channel:VelocityRatio}, we employ the quality criterion proposed by Rajkumar \etal \cite{Rajkumar2018} in this work, which can be calculated on each patch according to
\begin{equation}
    \label{eq:Channel:QualityCriterion}
	q_{\text{out},i} = \frac{\omega_{\text{out},i}-1}{\max(\omega_{\text{out},i}, 1)} \eqdot
\end{equation} 
\cref{eq:Channel:VelocityRatio,eq:Channel:QualityCriterion} confirm the previous argument that a block profile would be optimal with respect to the flow-homogeneity criterion: In case of a block profile, the average velocity on each patch would be identical to the average velocity of the whole outflow boundary, i.e., $v_{\text{out},i}=v_\text{avg}$, which in turn implies $\omega_{\text{out},i}=1$. With this, we obtain the theoretically optimal value of $q_{\text{out},i}=0$.

The patch-wise quality criteria $q_{\text{out}, i}$ are computed from the results of the \gls{CFD} simulation and then provided to the agent as observations. Moreover, they are used to determine the reward. The reward functions for both incremental and direct \gls{RL}-based shape optimizations are presented in the following section.


\subsection{Reward shaping}
\label{subsec:Channel:RewardShaping}

Similar to the test case with the T-shaped geometry presented in \cref{sec:TJunction}, we also shaped two reward functions for the direct and incremental approach. Since we used three patches in our geometry (see \cref{fig:Channel:OutflowPatches}) to compute the flow homogeneity at the outflow, we need to combine them into one single value $q_t$, which is done by employing a sum of squares
\begin{align}
\label{eq:Channel:QualitySum}
    q_t &= \sum_{i} {\left(q_{\text{out},i}\right)_t}^2 
    \eqdot
\end{align}
This single value is now used to determine the reward. Note that the block profile is also still optimal when considering \cref{eq:Channel:QualitySum}, since $q_{\text{out},i}=0$ also implies $q_t=0$.

\subsubsection{Direct Approach}
\label{subsec:Channel:RewardShaping:Direct}

The reward function for the direct optimization strategy is defined similar to the previous test case presented in \cref{eq:TJunction:Reward:Direct}
\begin{align}
\label{eq:Channel:Reward:Direct}
    r_t &= \begin{cases}
        -10 &  \text{simulation failed} \\
        -q_t & q_t \geq q^\star \\
        5 & q_t < q^\star
    \end{cases}
    \eqdot
\end{align} 
Unless the agent reaches the goal or the solver fails, the reward is proportional to the quality criterion and negative since the optimal quality criterion is $0$. As the direct optimization method's goal is to adapt the shape in a single step, each step where the agent fails to solve its task is undesirable. To be guided towards an optimal geometry, the agent receives a less severe penalty the closer the objective is to its goal.


\subsubsection{Incremental Approach}
\label{subsec:Channel:RewardShaping:Incremental}

In the incremental approach, the reward is defined based on the improvement between two succeeding steps $t-1$ and $t$ which reads
\begin{align}
\label{eq:Channel:Improvement}
    I_t &= q_{t-1} - q_{t} \eqdot
\end{align}
Based on that, the reward function is defined as
\begin{align}
\label{eq:Channel:Reward:Incremental}
    r_t = \begin{cases}
        -10 & \text{simulation failed} \\
        2I_t & I_t < 0 \\
        I_t & I_t \ge 0 \\
        5 & q_t < q^\star
    \end{cases}
    \eqdot 
\end{align} 
Here, the improvement $I_t$ is used as the reward if the improvement is greater than zero, meaning that the homogeneity improved compared to the last step. When the homogeneity deteriorates, $I_t$ is negative and again used as reward, but this time multiplied by a factor of two. This hinders the agent in exploiting the reward function and discourages it from undertaking actions that decrease the flow homogeneity. Once again, the reward for achieving the goal is $5$ and for tangling the mesh is $-10$.


\subsection{Agent Comparison}
\label{subsec:Channel:AgentComparison}

Similar to the previous test case, we start our investigation by comparing the different algorithms and examining their impact on training progression and stability. The experiments are shown for a limited number of steps, which is chosen such that all necessary information is contained in the selected interval.\footnote{The complete dataset can be provided by the authors upon request.} As before, all experiments in this section have been conducted twice. We first compare the agents trained with the direct shape optimization method, before comparing agents following an incremental shape optimization approach.


\subsubsection{Direct Approach}
\label{subsec:Channel:AgentComparison:Direct}

\begin{figure}[ht]
    \centering
    \input{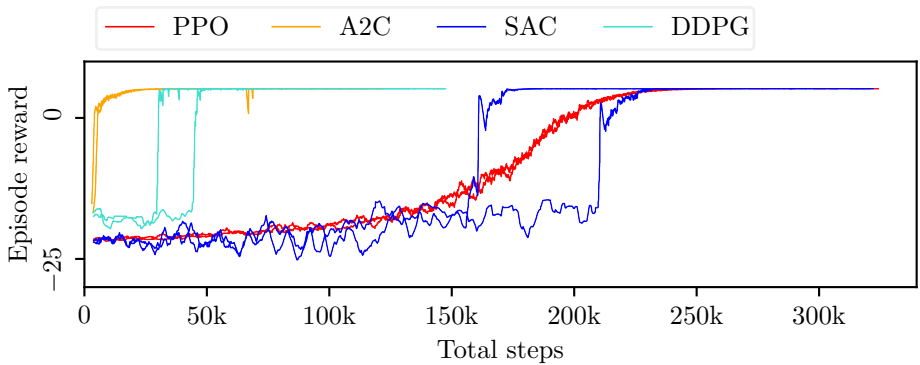}
    \caption{Comparison of different agents trained to optimize the converging channel geometry following a direct strategy with respect to the episode reward over the trained steps. Each run was repeated twice as indicated using the same color.}
    \label{fig:Channel:AgentComparison:Direct:EpisodeReward}
\end{figure}

\cref{fig:Channel:AgentComparison:Direct:EpisodeReward} depicts the agents' episode rewards over the total amount of trained steps. All results are shown for up to \SI{350}{\kilo{}} steps and at most \SI{100}{\kilo{}} episodes.
As a first observation, we note that for this test case, all agents manage to learn a strategy, as all reward curves converge to the optimal reward of $5$. This is in contrast to the previously presented optimization of the T-shaped geometry, where the \gls{A2C} agent did not learn anything. Secondly, we discover that the general shape of the agents' learning curves is similar between different runs, i.e., the algorithms seem to find the same strategies. However, only the learning curves of \gls{PPO} and \gls{A2C} are in very good agreement in terms of reproducibility. The sudden jump in the episode reward of the \gls{SAC} agent that we observed in \cref{fig:TJunction:AgentComparison:Direct:Steps} is also present in this test case. Yet, we note an identical behavior for the \gls{DDPG} agent, which did not show this behavior before. While converging quickly and monotonously to the maximal reward limit in the previous example, the \gls{DDPG} now behaves differently, displaying a behavior similar to the \gls{SAC} agent. However, the sharp jumps occur roughly \SI{100}{\kilo{}} steps earlier. The authors cannot definitely say, why the behaviour of this algorithm differs so much between the two examples. One explanation might be that the preceding test case was much simpler to learn as can be seen by the fewer steps required to learn a solution strategy. Therefore, the first phase of roughly constant episode reward, visible in \cref{fig:Channel:AgentComparison:Direct:EpisodeReward} for $t\leq \SI{30}{\kilo{}}$, is simply not present in \cref{fig:TJunction:AgentComparison:Direct:Steps}, as the agent directly found a way to increase the reward without first exploring the action space. 
One can also see that the \gls{A2C} and \gls{DDPG} agents are not trained for the whole \SI{300}{\kilo{}} steps. Instead, their training is stopped earlier when they reach the episode limit of \SI{100}{\kilo{}} episodes. This limit is reached so early since the agents are trained with a direct optimization approach: After being fully converged (i.e., $t\geq\SI{50}{\kilo{}}$), they only need a very small number of steps (often only one) per episode to accomplish their task. This can be seen in \cref{fig:Channel:OptimizationApproach:Direct}, where the steps per episode are plotted over the total number of training steps.

\begin{wraptable}{r}{5cm} 
    \centering
    \caption{Wall-clock times of the agents trained with the direct optimization method for the converging channel use case.}
    \begin{tabular}{lcc}
        \toprule
         Agent & Training time \\
         \midrule
         \gls{PPO} & 64.2 \si{\hour} \\
         \gls{A2C} & 44.0 \si{\hour} \\
         \gls{SAC} & 70.0 \si{\hour} \\
         \gls{DDPG} & 45.0 \si{\hour} \\
         \bottomrule
    \end{tabular}
    \label{tab:Channel:AgentComparison:Direct:WallTimes}
\end{wraptable}

Next, we investigate the performance of the different algorithms with respect to overall training time. Their wall-clock times differ widely as shown in \cref{tab:Channel:AgentComparison:Direct:WallTimes}. They range from approximately \SI{44}{\hour} for the \gls{A2C} and \gls{DDPG} agent to approximately \SI{70}{\hour} for the \gls{SAC} agent. One reason for these huge deviations is that the \gls{A2C} and \gls{DDPG} agents performed far fewer steps than the \gls{PPO} and \gls{SAC} agents as they managed to learn a strategy faster (as will later on be seen in \cref{fig:Channel:OptimizationApproach:Direct}). The difference between the \SI{64}{\hour} for the \gls{PPO} agent and the \SI{70}{\hour} for the \gls{SAC} agent shows that the former is more time efficient during the training. 
A more in-depth analysis of the differences with regard to the training time is complicated for the direct approach: Before starting a new episode, the environment needs to be reset, which involves a constant computational overhead that does not count toward the actual training time. However, as soon as the agent finds a good strategy, the number of steps per episode decreases significantly, in turn proportionally increasing the number of episode resets. When the agents manage to optimize the geometry in a single step, the time used for training is basically identical to the time for resetting the environment, complicating the task of determining the actual learning time.


\subsubsection{Incremental Approach}
\label{subsec:Channel:AgentComparison:Incremental}

\begin{figure}[ht]
    \centering
    \input{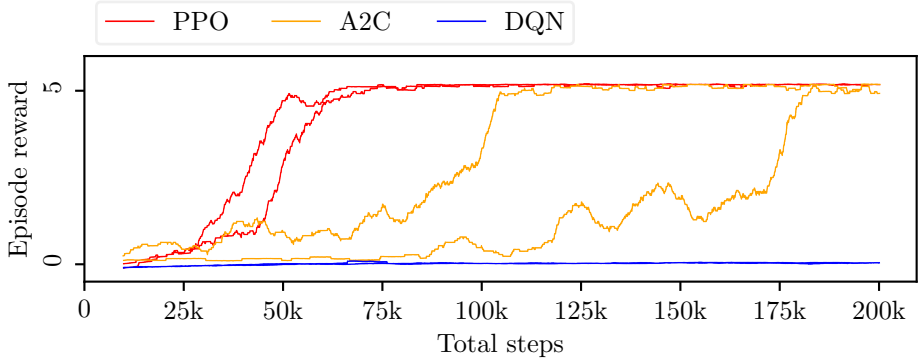}
    \caption{Comparison of different agents trained to optimize the converging channel geometry following an incremental strategy with respect to the episode reward over the trained steps. Each run was repeated twice as indicated using the same color.}
    \label{fig:Channel:AgentComparison:Incremental:EpisodeReward}
\end{figure}

Next, we compare the algorithms that are compatible with an incremental shape optimization approach. The achieved episode reward of the agents is shown for up to \SI{200}{\kilo{}} steps in \cref{fig:Channel:AgentComparison:Incremental:EpisodeReward}. 
In contrast to the results seen for the T-shaped geometry in \cref{fig:TJunction:AgentComparison:Incremental:Steps}, the \gls{A2C} agent manages to fully converge within the depicted number of steps, while the \gls{DQN} agent does not converge at all. It seems to initially make progress during the first \SI{100}{\kilo{}} steps but stagnates after that. The authors are not sure about the reason for this stagnation. It might be attributed to a not-optimal reward function, which seems to only work well for the \gls{PPO} agent as this algorithm repeatedly converges within $t\leq\SI{75}{\kilo{}}$ steps. However, different explanations are possible, e.g., the training progress of the \gls{DQN} agent might be generally much slower compared to the other algorithms (as it also only converges slowly when optimizing the T-shaped geometry as seen in \cref{fig:TJunction:AgentComparison:Incremental:Steps}) or the problem might simply be too complex for the agent to learn, given the employed combination of reward and observations. In terms of consistency between repeated training runs, \gls{A2C} now exhibits significant deviations. Clearly, further experiments are necessary to explain the behavior of the agents.

\begin{wraptable}{r}{5cm} 
    \centering
    \caption{Wall-clock times of the agents trained with the incremental optimization method for the converging channel use case.}
    \begin{tabular}{lcc}
        \toprule
         Agent & Training time \\
         \midrule
         \gls{PPO} & 42.2 \si{\hour} \\
         \gls{A2C} & 43.1 \si{\hour} \\
         \gls{DQN} & 44.7 \si{\hour} \\
         \bottomrule
    \end{tabular}
    \label{tab:Channel:AgentComparison:Incremental:WallTimes}
\end{wraptable}

The learning curves of the \gls{PPO} and \gls{A2C} agents increase smoother than the learning curves seen for the \gls{SAC} and \gls{DDPG} agents in the previous section, but less gradually than the \gls{PPO} agent trained on the T-shaped geometry following the same strategy.

As reported in \cref{tab:Channel:AgentComparison:Incremental:WallTimes}, the wall-clock times of the different algorithms are rather similar as opposed to the training times of the previous experiment. The difference between the \gls{DQN} agent and the \gls{PPO} agent only amounts to roughly \SI{2.5}{\hour}. This can partly be attributed to the fact that -- in the case of incremental shape optimization -- all agents train for the same number of steps and that they are not yet fully converged, as we will discuss in the next \cref{subsec:Channel:OptimizationApproach}.

Examples of optimized geometries obtained from a \gls{PPO} agent following an incremental shape optimization approach are included in \cref{app:Appendix:OptimizedGeometry:Channel}.


\subsection{Comparison of the optimization approaches: Direct vs. incremental shape optimization}
\label{subsec:Channel:OptimizationApproach}

\begin{figure}[ht]
    \centering
    \input{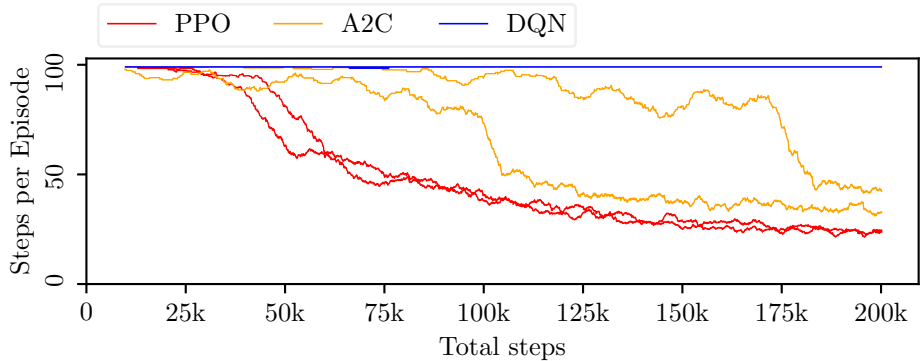}
    \caption{Comparison of different agents trained to optimize the converging channel geometry following an incremental strategy with respect to the steps per episode over the trained steps. Each run was repeated twice as indicated using the same color.}
    \label{fig:Channel:OptimizationApproach:Incremental}
\end{figure}

After investigating both \gls{RL}-based shape optimization approaches as introduced in \cref{subsec:Realisation:Actions} on two different examples, we want to use this section to compare both methods with the aim of providing an answer to \cref{itm:RQ1}.
\begin{figure}[ht]
    \centering
    \input{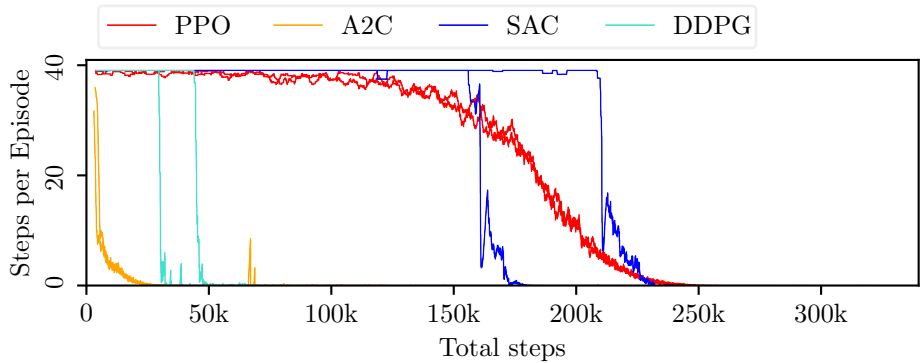}
    \caption{Comparison of different agents trained to optimize the converging channel geometry following a direct strategy with respect to the steps per episode over the trained steps. Each run was repeated twice as indicated using the same color.}
    \label{fig:Channel:OptimizationApproach:Direct}
\end{figure}
We have seen that different agents following either strategy can be successfully trained for both test cases. 
To validate that the strategies actually do what they promise, specifically that the direct optimization method is able to find an optimal geometry in only a single step, the number of training steps per episode (corresponding to one optimization each) has been analyzed and is depicted in \cref{fig:Channel:OptimizationApproach:Direct,fig:Channel:OptimizationApproach:Incremental} for both approaches. The difference between the results is apparent by comparing the $y$-axes and corresponds to our prior expectation. Once converged, agents following a direct optimization strategy decrease the required steps per episode significantly to -- on average -- one step per episode. In contrast, agents pursuing an incremental approach still need more than $30$ steps per episode, however, starting from initially $100$ steps, which were selected as a preliminary termination criterion.
In the plot for the incremental shape optimization strategy, we can also see that the algorithms are indeed not yet fully converged since they are still gradually decreasing the required steps per episode. 

This shows that both strategies can be successfully employed to solve the posed shape optimization problems. However, a clear statement of which optimization approach is generally preferable for our application cannot be made despite the extensive numerical experiments performed in this paper. This becomes especially obvious when comparing the \gls{PPO} and \gls{A2C} agent: When using a \gls{PPO} agent, it generally converges faster with the incremental strategy than with the direct optimization approach. This, however, does not hold true for the \gls{A2C} agent, where the behavior greatly varies depending on the problem under consideration.


\subsection{Vectorized Environment Training}
\label{subsec:Channel:MultiEnv}

\begin{figure}[ht]
    \centering
    \input{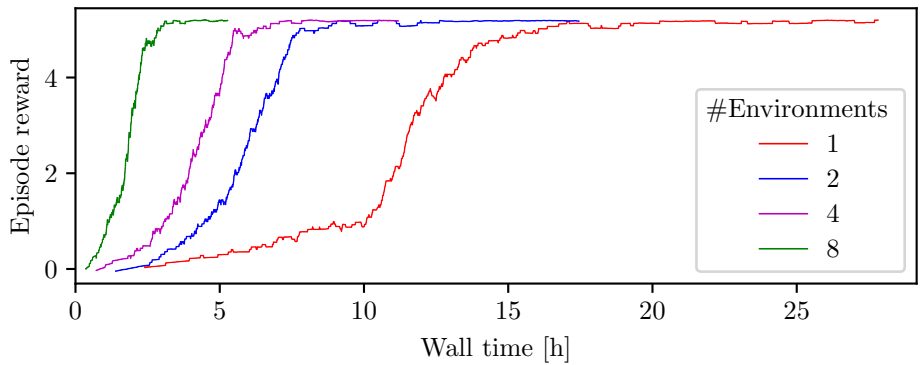}
    \caption{Episode reward over wall time of a \gls{PPO} agent following an incremental strategy for optimizing the converging channel geometry when interacting with different numbers of environments.}
    \label{fig:Channel:MultiEnv:WallTime}
\end{figure}

After investigating the benefit of utilizing vectorized environments on the T-shaped geometry, we also test it here for the converging channel test case. As seen in the previous sections, the \gls{PPO} agent trained with an incremental strategy once again proves to be relatively stable during training and is therefore again chosen for conducting these experiments. Every run trains a \gls{PPO} agent for \SI{125}{\kilo{}} steps since it is assumed that the \gls{PPO} agent should be converged in that interval (which seems a reasonable assumption, given the learning curve in \cref{fig:Channel:AgentComparison:Incremental:EpisodeReward}). 
\begin{figure}[ht]
    \centering
    \input{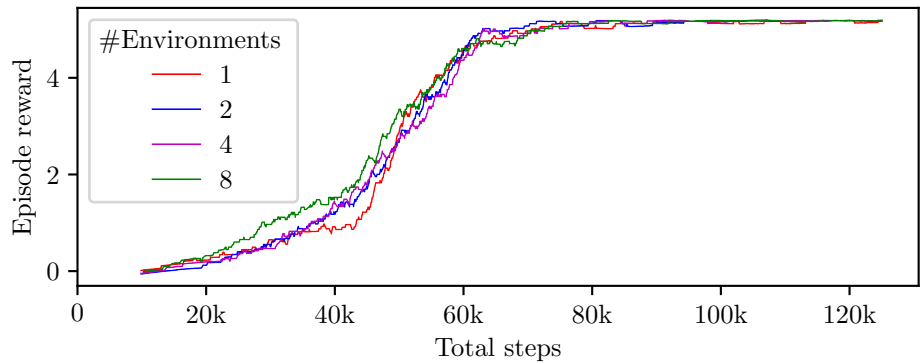}
    \caption{Episode reward over training steps of a \gls{PPO} agent following an incremental strategy for optimizing the converging channel geometry when interacting with different numbers of environments.}
    \label{fig:Channel:MultiEnv:Steps}
\end{figure}
The results of the vectorized environment training for the converging channel test case are presented in \cref{fig:Channel:MultiEnv:Steps} and generally show the same tendency as seen for the T-shaped geometry in \cref{fig:TJunction:MultiEnv:Steps}. The training progress made per step is once again very similar between the different experiments. When comparing the wall times of the different training runs, we observe a significant reduction when utilizing multiple vectorized environments as shown in \cref{fig:Channel:MultiEnv:WallTime}. Using eight environments, the wall time was reduced from roughly \SI{28}{\hour} to roughly \SI{5}{\hour}, corresponding to a reduction of roughly \SI{82}{\percent}.
With this, we can conclude that the utilization of vectorized environments is use case independent and, thus, an important tool in the fast iterative development of new application cases. 

It should be noted that the vectorized environment training did not scale completely linearly for our application. This means that despite a significant decrease in wall time, the used core hours\footnote{accumulated time consumed by all processor cores utilized in the computation} increase disproportionately with the utilized environments. This can partly be attributed to the increased overhead as well as to communication bottlenecks, i.e., some environments might need to wait for new instructions from the agent while it communicates with another environment or updates its policy. In addition, the scaling between both experiments with eight vectorized environments, i.e., comparing the scaling obtained for the T-shaped geometry to the scaling of the converging channel, differs. Therefore, \cref{itm:RQ3} can be answered qualitatively, while a quantitative answer needs further investigation.  
\section{Conclusion}
\label{sec:Conclusion}

Within this publication, we have presented an extensive investigation of different influence factors on the learning behavior of \gls{RL} learning algorithms for optimizing flow channels in profile extrusion dies. All our experiments have been conducted in a modular python-based framework, which combines the state-of-the-art \gls{RL} libraries \texttt{stable-baselines3} and \texttt{gym} with an \gls{FFD} implementation and our in-house \gls{FEM} solver for successfully training \gls{RL} agents for shape optimization tasks.

To answer \cref{itm:RQ1}, we have shown that \gls{RL} offers a promising perspective for repeatedly solving similar optimization tasks. In both our investigated application cases, a fully trained \gls{RL} agent using an incremental approach, choosing from $36$ discrete actions, can optimize a geometry in less than $50$ steps, whereas most likely even a gradient-based optimization algorithm would require more steps to obtain a solution. More extremely, an \gls{RL} agent trained following a direct strategy obtains an optimal geometry in a single step. Of course, the benefits of a trained agent need to be weighted against the time required for training. We cannot make a general statement, which of the presented approaches is preferable compared to the other, since only the implementations of two algorithms are compatible with both strategies and thus can be compared directly. Here, however, we found that training the agent to optimize a geometry directly took longer than training the same agent to find a good policy to incrementally optimize a geometry.

While our first test case investigated a more academic example, allowing to visually assess the correctness of the produced geometries, the second test case presented a more realistic example, utilizing \gls{RL} to optimize profile extrusion die flow channels with respect to a state-of-the-art design criterion. Here, the difficulty lies in the generalization of the learning task, i.e., how to train the agent to learn a generic strategy. In the present work, this was realized by providing the agent with randomly perturbed geometries at the beginning of each episode, i.e., training the agent to optimize a variety of geometries that could be realized by the chosen geometry parameterization. For future research, the generalization of the training process can definitely be addressed further, i.e., by providing image-based observations that include the whole flow field instead of only the control points of the deformation spline alongside selected post-processed quantities.

To address the second research question \cref{itm:RQ2}, we have shown that given the right reward function, all tested algorithms were able to find a strategy for generating optimal shapes. While -- as mentioned before -- we cannot make a general statement about the performance of direct compared to incremental shape optimization, agents following a direct strategy overall exhibit a better convergence behavior. Among the employed algorithms, we found \gls{PPO} to be most reliable in terms of consistency among repeated runs and convergence behavior in general, i.e., \gls{PPO} always manages to learn an effective strategy. With respect to convergence, \gls{PPO} is only outperformed by \gls{DDPG}. This algorithm exhibits a steep learning curve for both application cases and converges in a training time comparable to that of the \gls{PPO} agent. Furthermore, it also shows good agreement among repeated runs. \gls{A2C} was the only agent to find a way for exploiting the shaped reward function when optimizing the T-shaped geometry with a direct approach. However, for the test case of the converging channel, it converged fastest with respect to training time, highlighting the importance of properly shaping the reward function. \gls{DQN} shows only a slow learning progression for the T-shaped geometry, but does not learn anything for the converging channel geometry, rendering it less attractive for our application. Similar to \gls{A2C}, adaptions to the utilized reward function might cure this poor convergence, but this lies beyond the scope of this work. The learning curves of all \gls{SAC} agents exhibit one sharp jump at distinctly different training steps, thus lacking confident reproducibility. This behavior might result from a not-optimal learning rate, as we currently only looked at the default hyperparameters of the agents as implemented in \texttt{stable-baseline3}. It would subject of further studies, how much improvement can be achieved by optimizing the hyperparameters of the agents beforehand.

Finally, with respect to \cref{itm:RQ3}, we showed that training in vectorized environments significantly reduces the wall-time. The approach proposed by Rabault \etal \cite{Rabault2019} successfully worked on both geometries, achieving speed-ups of ${\sim}4.6$ for the T-shaped geometry and even  ${\sim}5.6$ for the converging channel. Of course, the scaling of this approach might greatly depend on the framework under consideration, as well as the utilized algorithm and shape optimization approach. For this paper, we restricted ourselves to a \gls{PPO} agent, following an incremental shape optimization approach.

While we believe our findings to be a first step towards a more thorough understanding of the different factors impacting the learning behaviour of \gls{RL} algorithms, still plenty of tasks remain open, which need to be addressed in the future: Firstly, based on this work, a suitable algorithm and shape optimization approach might be selected to extend the presented approach to realistic 3D geometries, favorably with the possibility to test the optimized flow channel during operation. Here, it is also interesting to see if a trained agent can transfer its learned knowledge, i.e., it's policy, to slightly different geometries. Finally, as already alluded to above, one could investigate ways to provide more expressive observations to the agent, i.e., information from the whole flow field, to make the learning process more general and less dependent on the deformation spline used for parameterization.

\section*{Acknowledgments}
This work was partially performed as part of the Helmholtz School for Data Science in Life, Earth and Energy (HDS-LEE) and received funding from the Helmholtz Association of German Research Centres. 
Funded by the Deutsche Forschungsgemeinschaft (DFG, German Research Foundation) under Germany's Excellence Strategy – EXC-2023 Internet of Production – 390621612.
Simulations were performed with computing resources granted by RWTH Aachen University under projects \texttt{thes1136} and \texttt{jara0185}.









\bibliographystyle{AIMS}
\bibliography{literature.bib}

\providecommand{\href}[2]{#2}
\providecommand{\arxiv}[1]{\href{http://arxiv.org/abs/#1}{arXiv:#1}}
\providecommand{\url}[1]{\texttt{#1}}
\providecommand{\urlprefix}{URL }
\begin{thebibliography}{10}

\bibitem{Arulkumaran2017}
\newblock K.~Arulkumaran, M.~P. Deisenroth, M.~Brundage and A.~A. Bharath,
\newblock {Deep reinforcement learning: A brief survey},
\newblock \emph{IEEE Signal Processing Magazine}, \textbf{34} (2017), 26--38.

\bibitem{Brockman2016}
\newblock G.~Brockman, V.~Cheung, L.~Pettersson, J.~Schneider, J.~Schulman,
  J.~Tang and W.~Zaremba,
\newblock {OpenAI Gym},
\newblock 1--4,
\newblock \urlprefix\url{http://arxiv.org/abs/1606.01540}.

\bibitem{Busoniu2018}
\newblock L.~Buşoniu, T.~de~Bruin, D.~Toli{\'{c}}, J.~Kober and I.~Palunko,
\newblock {Reinforcement learning for control: Performance, stability, and deep
  approximators},
\newblock \emph{Annual Reviews in Control}, \textbf{46} (2018), 8--28.

\bibitem{Carreau1972}
\newblock P.~J. Carreau,
\newblock {Rheological Equations from Molecular Network Theories},
\newblock \emph{Transactions of the Society of Rheology}, \textbf{16} (1972),
  99--127,
\newblock \urlprefix\url{http://sor.scitation.org/doi/10.1122/1.549276}.

\bibitem{Cottrell2009}
\newblock J.~A. Cottrell, T.~J.~R. Hughes and Y.~Bazilevs,
\newblock \emph{Isogeometric Analysis},
\newblock John Wiley {\&} Sons, Ltd, Chichester, UK, 2009,
\newblock \urlprefix\url{http://doi.wiley.com/10.1002/9780470749081}.

\bibitem{Dworschak2022}
\newblock F.~Dworschak, S.~Dietze, M.~Wittmann, B.~Schleich and S.~Wartzack,
\newblock {Reinforcement Learning for Engineering Design Automation},
\newblock \emph{Advanced Engineering Informatics}, \textbf{52} (2022), 101612,
\newblock \urlprefix\url{https://doi.org/10.1016/j.aei.2022.101612}.

\bibitem{Elgeti2012}
\newblock S.~Elgeti, M.~Probst, C.~Windeck, M.~Behr, W.~Michaeli and
  C.~Hopmann,
\newblock {Numerical shape optimization as an approach to extrusion die
  design},
\newblock \emph{Finite Elements in Analysis and Design}, \textbf{61} (2012),
  35--43,
\newblock \urlprefix\url{http://dx.doi.org/10.1016/j.finel.2012.06.008}.

\bibitem{Garnier2021}
\newblock P.~Garnier, J.~Viquerat, J.~Rabault, A.~Larcher, A.~Kuhnle and
  E.~Hachem,
\newblock {A review on deep reinforcement learning for fluid mechanics},
\newblock \emph{Computers and Fluids}, \textbf{225}.

\bibitem{Ghraieb2022}
\newblock H.~Ghraieb, J.~Viquerat, A.~Larcher, P.~Meliga and E.~Hachem,
\newblock {Single-step deep reinforcement learning for two- and
  three-dimensional optimal shape design},
\newblock \emph{AIP Advances}, \textbf{12} (2022), 085108,
\newblock \urlprefix\url{https://doi.org/10.1063/5.0097241}.

\bibitem{Haarnoja2018}
\newblock T.~Haarnoja, A.~Zhou, P.~Abbeel and S.~Levine,
\newblock {Soft actor-critic: Off-policy maximum entropy deep reinforcement
  learning with a stochastic actor},
\newblock \emph{35th International Conference on Machine Learning, ICML 2018},
  \textbf{5} (2018), 2976--2989.

\bibitem{Hopmann2016}
\newblock C.~Hopmann and W.~Michaeli,
\newblock \emph{{Extrusion Dies for Plastics and Rubber}},
\newblock 4th edition,
\newblock Carl Hanser Verlag GmbH {\&} Co. KG, M{\"{u}}nchen, 2016,
\newblock
  \urlprefix\url{http://www.hanser-elibrary.com/doi/book/10.3139/9781569906248}.

\bibitem{Kaelbling1996}
\newblock L.~P. Kaelbling, M.~L. Littman and A.~W. Moore,
\newblock {Reinforcement Learning: A Survey},
\newblock \emph{Journal of Artificial Intelligence Research}, \textbf{4}
  (1996), 237--285,
\newblock \urlprefix\url{https://dl.acm.org/doi/10.5555/1622737.1622748
  http://arxiv.org/abs/cs/9605103}.

\bibitem{Kober2014}
\newblock J.~Kober and J.~Peters,
\newblock \emph{Reinforcement Learning in Robotics: A Survey}, 9--67,
\newblock Springer International Publishing, Cham, 2014,
\newblock \urlprefix\url{https://doi.org/10.1007/978-3-319-03194-1_2}.

\bibitem{Konda2000}
\newblock V.~R. Konda and J.~N. Tsitsiklis,
\newblock {Actor-critic algorithms},
\newblock \emph{Advances in Neural Information Processing Systems}, 1008--1014.

\bibitem{Lampton2008}
\newblock A.~Lampton, A.~Niksch and J.~Valasek,
\newblock {Morphing airfoils with four morphing parameters},
\newblock \emph{AIAA Guidance, Navigation and Control Conference and Exhibit}.

\bibitem{gustaf}
\newblock J.~Lee,
\newblock guastaf,
\newblock \urlprefix\url{https://github.com/tataratat/gustaf}.

\bibitem{Li2021}
\newblock R.~Li, Y.~Zhang and H.~Chen,
\newblock {Learning the Aerodynamic Design of Supercritical Airfoils Through
  Deep Reinforcement Learning},
\newblock \emph{AIAA Journal}, \textbf{59} (2021), 3988--4001.

\bibitem{Lillicrap2016}
\newblock T.~P. Lillicrap, J.~J. Hunt, A.~Pritzel, N.~Heess, T.~Erez, Y.~Tassa,
  D.~Silver and D.~Wierstra,
\newblock {Continuous control with deep reinforcement learning},
\newblock \emph{4th International Conference on Learning Representations, ICLR
  2016 - Conference Track Proceedings}.

\bibitem{Michaeli2001}
\newblock W.~Michaeli, S.~Kaul and T.~Wolff,
\newblock {COMPUTER-AIDED OPTIMIZATION OF EXTRUSION DIES},
\newblock \emph{Journal of Polymer Engineering}, \textbf{21},
\newblock
  \urlprefix\url{https://www.degruyter.com/document/doi/10.1515/POLYENG.2001.21.2-3.225/html}.

\bibitem{Mnih2016}
\newblock V.~Mnih, A.~P. Badia, L.~Mirza, A.~Graves, T.~Harley, T.~P.
  Lillicrap, D.~Silver and K.~Kavukcuoglu,
\newblock {Asynchronous methods for deep reinforcement learning},
\newblock \emph{33rd International Conference on Machine Learning, ICML 2016},
  \textbf{4} (2016), 2850--2869.

\bibitem{Mnih2013}
\newblock V.~Mnih, K.~Kavukcuoglu, D.~Silver, A.~Graves, I.~Antonoglou,
  D.~Wierstra and M.~Riedmiller,
\newblock {Playing Atari with Deep Reinforcement Learning},
\newblock 1--9,
\newblock \urlprefix\url{http://arxiv.org/abs/1312.5602}.

\bibitem{Nobrega2004}
\newblock J.~M. N{\'{o}}brega, O.~S. Carneiro, F.~T. Pinho and P.~J. Oliveira,
\newblock {Flow balancing in extrusion dies for thermoplastic profiles, Part
  III: Experimental assessment},
\newblock \emph{International Polymer Processing}, \textbf{19} (2004),
  225--235.

\bibitem{Osswald2015}
\newblock T.~Osswald and N.~Rudolph,
\newblock \emph{{Polymer Rheology}},
\newblock Carl Hanser Verlag GmbH {\&} Co. KG, 2015.

\bibitem{Pauli2013}
\newblock L.~Pauli, M.~Behr and S.~Elgeti,
\newblock {Towards shape optimization of profile extrusion dies with respect to
  homogeneous die swell},
\newblock \emph{Journal of Non-Newtonian Fluid Mechanics}, \textbf{200} (2013),
  79--87,
\newblock \urlprefix\url{http://dx.doi.org/10.1016/j.jnnfm.2012.12.002}.

\bibitem{Piegl1995}
\newblock L.~Piegl and W.~Tiller,
\newblock \emph{The NURBS Book},
\newblock Monographs in Visual Communications, Springer Berlin Heidelberg,
  Berlin, Heidelberg, 1995,
\newblock \urlprefix\url{http://link.springer.com/10.1007/978-3-642-97385-7}.

\bibitem{Pittman2011}
\newblock J.~F. Pittman,
\newblock {Computer-aided design and optimization of profile extrusion dies for
  thermoplastics and rubber: A review},
\newblock \emph{Proceedings of the Institution of Mechanical Engineers, Part E:
  Journal of Process Mechanical Engineering}, \textbf{225} (2011), 280--321.

\bibitem{Qin2021}
\newblock S.~Qin, S.~Wang, L.~Wang, C.~Wang, G.~Sun and Y.~Zhong,
\newblock {Multi-objective optimization of cascade blade profile based on
  reinforcement learning},
\newblock \emph{Applied Sciences (Switzerland)}, \textbf{11} (2021), 1--27.

\bibitem{Rabault2019}
\newblock J.~Rabault and A.~Kuhnle,
\newblock Accelerating deep reinforcement learning strategies of flow control
  through a multi-environment approach,
\newblock \emph{Physics of Fluids}, \textbf{31}.

\bibitem{Rabault2020}
\newblock J.~Rabault, F.~Ren, W.~Zhang, H.~Tang and H.~Xu,
\newblock {Deep reinforcement learning in fluid mechanics: A promising method
  for both active flow control and shape optimization},
\newblock \emph{Journal of Hydrodynamics}, \textbf{32} (2020), 234--246.

\bibitem{Raffin2021}
\newblock A.~Raffin, A.~Hill, A.~Gleave, A.~Kanervisto, M.~Ernestus and
  N.~Dormann,
\newblock {Stable-baselines3: Reliable reinforcement learning implementations},
\newblock \emph{Journal of Machine Learning Research}, \textbf{22} (2021),
  1--8.

\bibitem{Rajkumar2018}
\newblock A.~Rajkumar, L.~L. Ferr{\'{a}}s, C.~Fernandes, O.~S. Carneiro and
  J.~M. N{\'{o}}brega,
\newblock {Guidelines for balancing the flow in extrusion dies: The influence
  of the material rheology},
\newblock \emph{Journal of Polymer Engineering}, \textbf{38} (2018), 197--211.

\bibitem{Rajkumar2017}
\newblock A.~Rajkumar, L.~L. Ferrás, C.~Fernandes, O.~S. Carneiro, M.~Becker
  and J.~M. Nóbrega,
\newblock Design guidelines to balance the flow distribution in complex profile
  extrusion dies,
\newblock \emph{International Polymer Processing}, \textbf{32} (2017), 58--71,
\newblock \urlprefix\url{https://doi.org/10.3139/217.3272}.

\bibitem{Schulman2017}
\newblock J.~Schulman, F.~Wolski, P.~Dhariwal, A.~Radford and O.~Klimov,
\newblock {Proximal Policy Optimization Algorithms},
\newblock \emph{CoRR}, 1--12,
\newblock \urlprefix\url{http://arxiv.org/abs/1707.06347}.

\bibitem{Sederberg1986}
\newblock T.~W. Sederberg and S.~R. Parry,
\newblock {Free-form deformation of solid geometric models},
\newblock \emph{Proceedings of the 13th Annual Conference on Computer Graphics
  and Interactive Techniques, SIGGRAPH 1986}, \textbf{20} (1986), 151--160.

\bibitem{Siegbert2014}
\newblock R.~Siegbert, J.~Kitschke, H.~Djelassi, M.~Behr and S.~Elgeti,
\newblock {Comparing Optimization Algorithms for Shape Optimization of
  Extrusion Dies},
\newblock \emph{PAMM}, \textbf{14} (2014), 789--794,
\newblock
  \urlprefix\url{https://onlinelibrary.wiley.com/doi/10.1002/pamm.201410377}.

\bibitem{Silver2018}
\newblock D.~Silver, T.~Hubert, J.~Schrittwieser, I.~Antonoglou, M.~Lai,
  A.~Guez, M.~Lanctot, L.~Sifre, D.~Kumaran, T.~Graepel, T.~Lillicrap,
  K.~Simonyan and D.~Hassabis,
\newblock {A general reinforcement learning algorithm that masters chess,
  shogi, and Go through self-play},
\newblock \emph{Science}, \textbf{362} (2018), 1140--1144.

\bibitem{Sutton2018}
\newblock R.~S. Sutton and A.~G. Barto,
\newblock \emph{Reinforcement Learning: An Introduction},
\newblock 2nd edition,
\newblock The MIT Press, 2018,
\newblock \urlprefix\url{http://incompleteideas.net/book/the-book-2nd.html}.

\bibitem{Szarvasy2000}
\newblock I.~Szarvasy, J.~Sienz, J.~F.~T. Pittman and E.~Hinton,
\newblock {Computer Aided Optimisation of Profile Extrusion Dies},
\newblock \emph{International Polymer Processing}, \textbf{15} (2000), 28--39.

\bibitem{Tezduyar1992a}
\newblock T.~E. Tezduyar, J.~Liou and M.~Behr,
\newblock {A new strategy for finite element computations involving moving
  boundaries and interfaces--the DSD/ST procedure: I. The concept and the
  preliminary numerical tests},
\newblock \emph{Computer Methods in Applied Mechanics and Engineering},
  \textbf{94} (1992), 339--351.

\bibitem{Vinyals2019}
\newblock O.~Vinyals, I.~Babuschkin, W.~M. Czarnecki, M.~Mathieu, A.~Dudzik,
  J.~Chung, D.~H. Choi, R.~Powell, T.~Ewalds, P.~Georgiev, J.~Oh, D.~Horgan,
  M.~Kroiss, I.~Danihelka, A.~Huang, L.~Sifre, T.~Cai, J.~P. Agapiou,
  M.~Jaderberg, A.~S. Vezhnevets, R.~Leblond, T.~Pohlen, V.~Dalibard,
  D.~Budden, Y.~Sulsky, J.~Molloy, T.~L. Paine, C.~Gulcehre, Z.~Wang, T.~Pfaff,
  Y.~Wu, R.~Ring, D.~Yogatama, D.~W{\"{u}}nsch, K.~McKinney, O.~Smith,
  T.~Schaul, T.~Lillicrap, K.~Kavukcuoglu, D.~Hassabis, C.~Apps and D.~Silver,
\newblock {Grandmaster level in StarCraft II using multi-agent reinforcement
  learning},
\newblock \emph{Nature}, \textbf{575} (2019), 350--354,
\newblock \urlprefix\url{http://dx.doi.org/10.1038/s41586-019-1724-z}.

\bibitem{Viquerat2021PBO}
\newblock J.~Viquerat, R.~Duvigneau, P.~Meliga, A.~Kuhnle and E.~Hachem,
\newblock {Policy-based optimization: single-step policy gradient method seen
  as an evolution strategy},
\newblock \urlprefix\url{http://arxiv.org/abs/2104.06175}.

\bibitem{Viquerat2021Review}
\newblock J.~Viquerat, P.~Meliga and E.~Hachem,
\newblock {A review on deep reinforcement learning for fluid mechanics: an
  update},
\newblock \urlprefix\url{http://arxiv.org/abs/2107.12206}.

\bibitem{Viquerat2021Paper}
\newblock J.~Viquerat, J.~Rabault, A.~Kuhnle, H.~Ghraieb, A.~Larcher and
  E.~Hachem,
\newblock {Direct shape optimization through deep reinforcement learning},
\newblock \emph{Journal of Computational Physics}, \textbf{428}.

\bibitem{Wolff2022}
\newblock D.~Wolff, C.~D. Fricke, M.~Kemmerling and S.~Elgeti,
\newblock {[WIP] Towards shape optimization of flow channels in profile
  extrusion dies using reinforcement learning},
\newblock \emph{Proceedings in Applied Mathematics and Mechanics}, \textbf{22}.

\bibitem{Yan2019}
\newblock X.~Yan, J.~Zhu, M.~Kuang and X.~Wang,
\newblock {Aerodynamic shape optimization using a novel optimizer based on
  machine learning techniques},
\newblock \emph{Aerospace Science and Technology}, \textbf{86} (2019),
  826--835,
\newblock \urlprefix\url{https://doi.org/10.1016/j.ast.2019.02.003}.

\bibitem{Yilmaz2014}
\newblock O.~Yilmaz, H.~Gunes and K.~Kirkkopru,
\newblock {Optimization of a profile extrusion die for flow balance},
\newblock \emph{Fibers and Polymers}, \textbf{15} (2014), 753--761.

\bibitem{Zhang2019a}
\newblock G.~Zhang, X.~Huang, S.~Li and T.~Deng,
\newblock {Optimized Design Method for Profile Extrusion Die Based on NURBS
  Modeling},
\newblock \emph{Fibers and Polymers}, \textbf{20} (2019), 1733--1741.

\end{thebibliography}

\medskip
Received xxxx 20xx; revised xxxx 20xx; early access xxxx 20xx.
\medskip

\clearpage

\section{Appendix}
\label{app:Appendix}

The appendix includes additional information and figures.

\subsection{Optimized Geometries}
This section contains examples of optimized geometries for both test cases presented in \cref{sec:TJunction,sec:Channel}.

\subsubsection{T-shaped geometry}
\label{app:Appendix:OptimizedGeometry:TJunction}

Results are obtained with a \gls{PPO} agent trained to follow the incremental strategy as described in \cref{subsec:Realisation:Actions} for the T-shaped geometry presented in \cref{sec:TJunction}.\newline

\begin{figure}[ht]
    \centering

    \subfloat[Geometry with goal ratio $\mu^\star=0.2$.]{\includegraphics[width=0.45\linewidth]{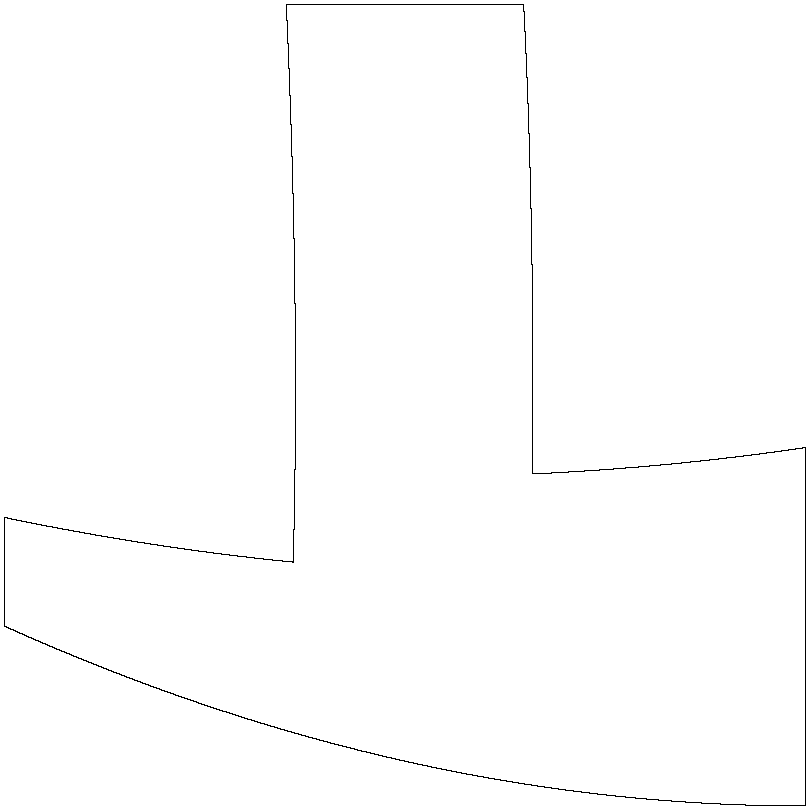}}
    \hspace{0.025\linewidth}
    \subfloat[Geometry with goal ratio $\mu^\star=0.6$.]{\includegraphics[width=0.45\linewidth]{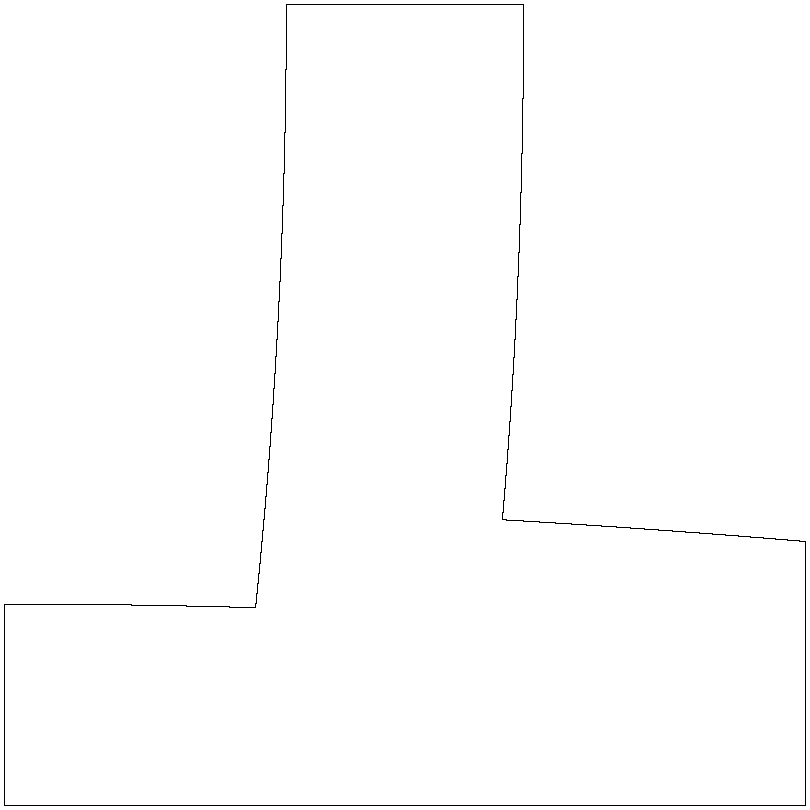}}
    
    \subfloat[Geometry with goal ratio $\mu^\star=0.9$.]{\includegraphics[width=0.45\linewidth]{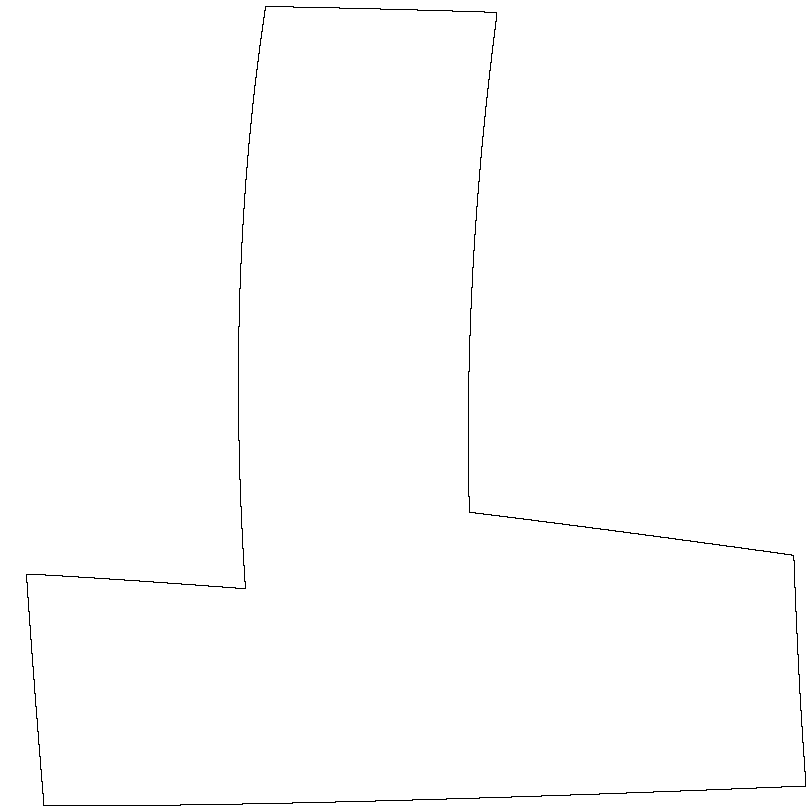}}
    \hspace{0.025\linewidth}
    \subfloat[Geometry with goal ratio $\mu^\star=1.4$.]{\includegraphics[width=0.45\linewidth]{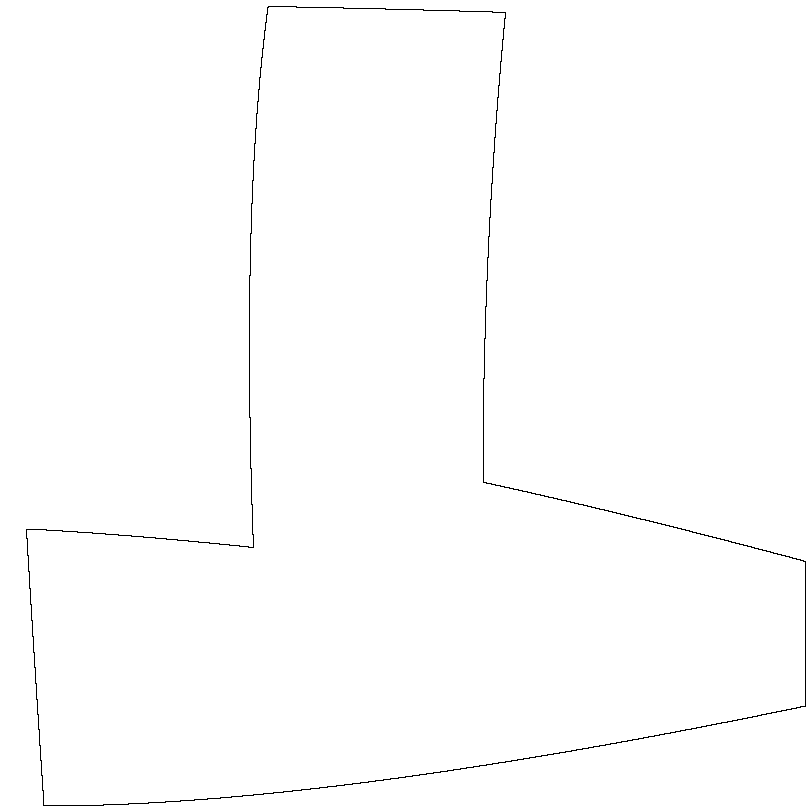}}
    
    \caption{Examples of optimal geometries obtained by a \gls{PPO} agent for the use-case with the T-shaped geometry following an incremental optimization strategy.}
    \label{fig:Appendix:OptimizedGeometry:TJunction}
\end{figure}

\clearpage


\subsubsection{Converging Channel}
\label{app:Appendix:OptimizedGeometry:Channel}
\vspace{-0.2cm}
Results are obtained with a \gls{PPO} agent trained to follow the incremental strategy as described in \cref{subsec:Realisation:Actions} for the converging channel test case presented in \cref{sec:Channel}. Since each episode is initialized with a different, randomly generated channel geometry, each subfigure depicts the optimized geometry of that episode.\newline

\begin{figure}[ht]
    \centering
    \subfloat[Optimized geometry of episode $274$.]{\includegraphics[width=0.45\linewidth]{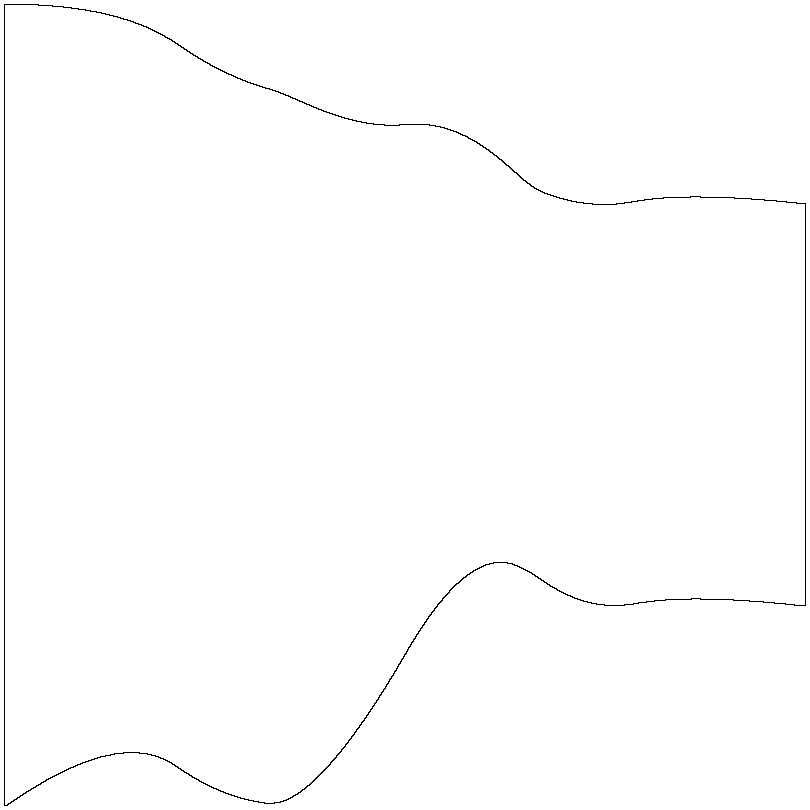}}
    \hspace{0.025\linewidth}
    \subfloat[Optimized geometry of episode $1760$.]{\includegraphics[width=0.45\linewidth]{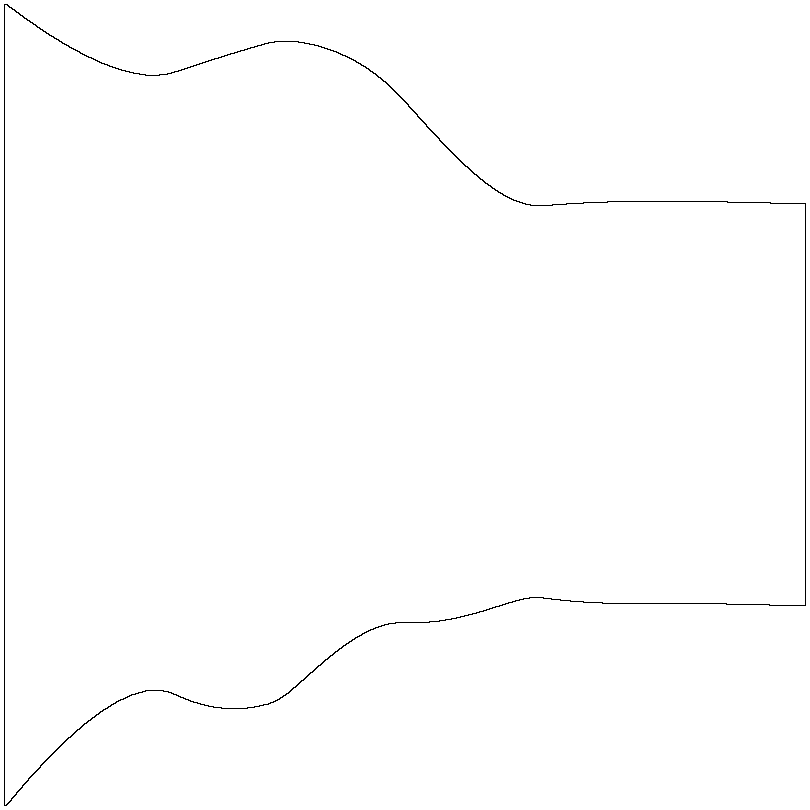}}
    
    \subfloat[Optimized geometry of episode $3127$.]{\includegraphics[width=0.45\linewidth]{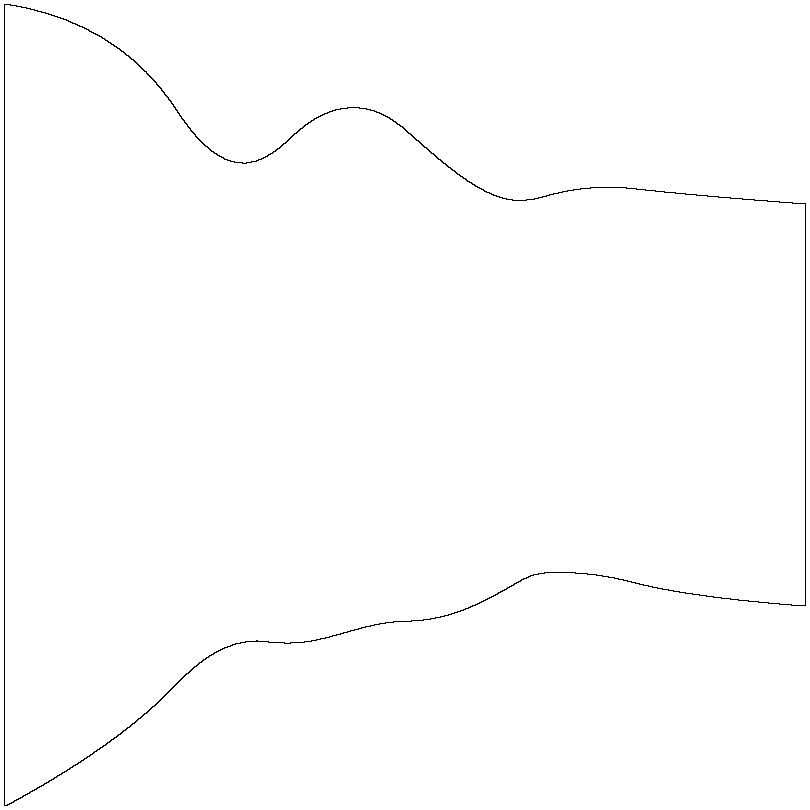}}
    \hspace{0.025\linewidth}
    \subfloat[Optimized geometry of episode $3597$.]{\includegraphics[width=0.45\linewidth]{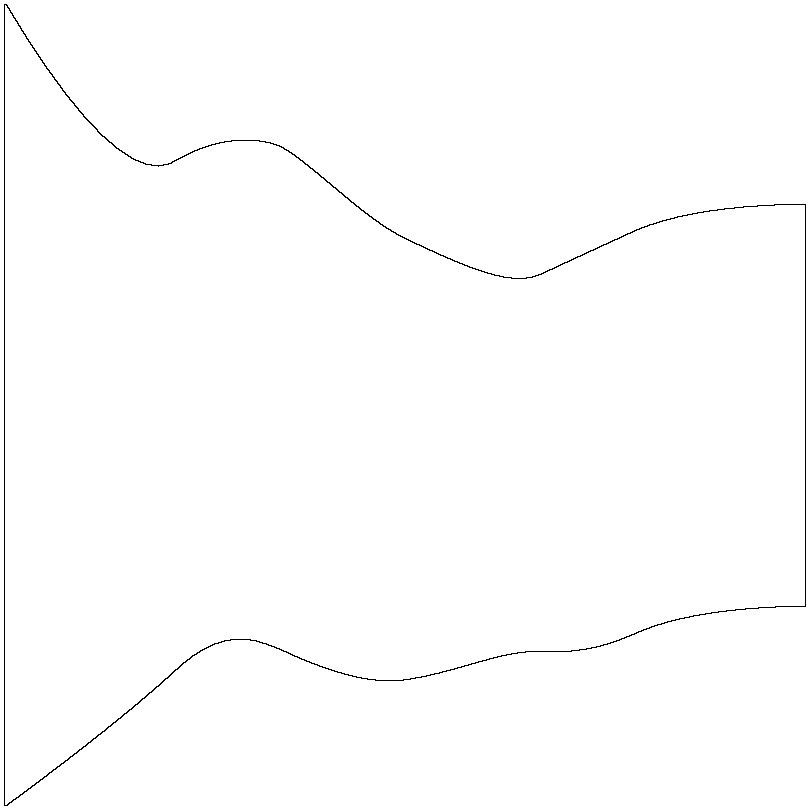}}
    \caption{Examples of optimal geometries obtained by a \gls{PPO} agent for the converging channel use-case following an incremental optimization strategy.}
    \label{fig:Appendix:OptimizedGeometry:Channel}
\end{figure}

\end{document}